\documentclass[aps,pre,preprint,groupedaddress,amsmath,amssymb,linenumbers]{revtex4}
\usepackage[utf8]{inputenc}
\usepackage[english]{babel}
\newtheorem{theorem}{Theorem}

\usepackage{epsfig,amsmath,amssymb,bm,epsf,graphics,graphicx,psfrag,verbatim,subfigure,framed}
\usepackage{float}
\usepackage{color}
\usepackage{url}
\usepackage{amsfonts}
\usepackage{mathrsfs}
\usepackage{indentfirst}
\DeclareMathAlphabet{\mathpzc}{OT1}{pzc}{m}{it}
\def\newblock{\hskip .11em plus .33em minus .07em}

\newcommand{\xv}{{\mathbf x}}

\newcommand{\qv}{{\mathbf q}}
\newcommand{\pv}{{\mathbf p}}
\newcommand{\fv}{{\bf f}}
\newcommand{\yv}{{\mathbf y}}

\newcommand{\be}{\begin{equation}}
\newcommand{\ee}{\end{equation}} 
\newcommand{\ba}{\begin{eqnarray}}
\newcommand{\ea}{\end{eqnarray}}

\begin{document}

%\preprint{APS/123-QED}
\title{Definitions and Evolutions of Statistical Entropy for  Hamiltonian Systems }
\author{Xiangjun Xing}
\email{xxing@sjtu.edu.cn}
\address{School of Physics and Astronomy, Shanghai Jiao Tong University, Shanghai, 200240 China
\\
Collaborative Innovation Center of Advanced Microstructures, Nanjing 210093, China}

%\affiliation{}
%Lines break automatically or can be forced with \\
\date{\today} %freeze this upon submission
% It is always \today, today,
% but any date may be explicitly specifie
%\pacs{82.70.Dd, 83.80.Hj, 82.45.Gj, 52.25.Kn}

% [CHECK: need to add PACS numbers]
%\pacs{61.41.+e, 61.43.Er, 62.20.Dc, 64.60.Ak, 64.60.Fr, 64.70.Dv}% PACS, the 
%61.41.+e 61.43.-j
%Physics and Astronomy
% Classification Scheme.
%\keywords{Suggested keywords}%Use showkeys class option if keyword
%display desired

\begin{abstract}
Regardless of studies and debates over a century, the true meaning of the second law of thermodynamics still remains illusive.  One outstanding question is whether the entropy of a closed system increases monotonically, or just probabilistically.  Here I revisit the seminal ideas about non-equilibrium statistical entropy due to Boltzmann and due to Gibbs, and synthesize them into a coherent and precise framework. Using this framework, I clarify the anthropomorphic principle of entropy, and analyze the evolution of entropy for classical Hamiltonian systems under different experimental setups.  I find that evolution of Boltzmann entropy obeys a {\em Stochastic H-Theorem}, which relates probability of Boltzmann entropy increasing to that of decreasing.  By contrast, the coarse-grained Gibbs entropy is monotonically increasing, if the microscopic dynamics is locally mixing, and the initial state is a Boltzmann state.  These results clarify the precise meaning of the second law of thermodynamics for classical systems, and demonstrate that it is the initial condition as a Boltzmann state that is ultimately responsible for the arrow of time.

\vspace{-2mm}

\end{abstract}

\maketitle

 Entropy is perhaps the most confusing concept in physics. According to the second law of  thermodynamics~\cite{Caratheodory-1909,Sommerfeld-book}, there exists for {\em closed systems} at equilibrium a state variable $S_{\rm T}$, called {\em thermodynamic entropy},  whose values are the same for states mutually accessible via reversible processes.  If two states are connected by an irreversible process, then $S_{\rm T}$ of the final state must be larger than that of the initial state.   It follows that the thermodynamic entropy achieves the maximally possible value in thermal equilibrium.  The second law says nothing about entropy in generic non-equilibrium states. 

% Nonetheless, the 

Gibbs~\cite{Gibbs-principle} defined a {\em statistical entropy} for an arbitrary phase space probability density function (pdf) $\rho(\xv)$ as
\vspace{-1mm}
\be
S_{\rm G}[\rho(\xv)] = {\displaystyle
 - \int_{\Omega} \rho({{\mathbf x}}) \log \rho({{\mathbf x}}) d {\mathbf x}} , 
\label{S-def-Gibbs}
\ee
which is usually called the {\em Gibbs entropy}.  This concept can be generalized to quantum systems in a straightforward fashion, as shown by von Neumann.  By maximizing $S_{\rm G}$ subject to relevant constraints, one can obtain various equilibrium ensembles as well as all equilibrium thermodynamic relations, and thereby establish the full correspondence between  thermodynamics and statistical mechanics in equilibrium.  This allows us to identify Gibbs entropy with thermodynamic entropy in equilibrium situations.  The correspondence is unfortunately invalid for non-equilibrium systems. According to the Liouville theorem~\cite{Landau-mechanics}, phase space volume is conserved by Hamiltonian dynamics.  This means that  $S_{\rm G}[\rho(\xv)]$ of a closed system is {\em conserved}, yet thermodynamics says that $S_{\rm T}$ increases monotonically.  Hence for non-equilibrium systems, these two concepts must be different.  To resolve this paradox, Gibbs~\cite{Ehrenfest,Gibbs-principle} argued that it is the Gibbs entropy of {\em coarse-grained} probability density distribution $\tilde{\rho}({\mathbf x})$, that should be identified with thermodynamic entropy.  However, it is not yet known whether the coarse-grained entropy increases monotonically over time.  The apparent arbitrariness in the choice of coarse-graining method is another itching issue~\cite{Jaynes-1965-entropy}.  % In summary,  before we can resolve the conflict between Liouville theorem and the second law, we do not know what entropy is for non-equilibrium systems.  

%Gibbs entropy Eq.~(\ref{S-def-Gibbs}) appears to be a {\em subjective concept}, because it measures the ignorance of the observer.  Maxwell invoked his famous demon who can measure details of the system and use them to lower the entropy, or to do useful work.  If the demon measure every single detail of a system, does it reduce the entropy down to zero?  But this entropy change reflects change of demon's knowledge, not of the system itself.  This paradox has inspired generations of physicists to explore the true meaning of entropy in non-equilibrium situations.  Additionally, Jaynes also challenged the objectivity of coarse-graining and the coarse-grained Gibbs entropy~\cite{Jaynes-1965-entropy}: ``$ \cdots $ The difference between H and $\tilde{H}$  is characteristic, not of the macroscopic state, but of the particular way in which we choose to coarse-grain. $\cdots$''. 

%The connection between the second law and Gibbs entropy remains to be convincingly established.  

%\xing{Many works followed Gibbs attempting to establish the connection.  NO general agreement. }

%Entropy seems to be as much perplexing as the wave function of quantum mechanics. 

{ Boltzmann} devoted his entire career to the statistical origin of entropy.  Using his kinetic equation, Boltzmann derived the H-theorem~\cite{Ehrenfest,Boltzmann-biography,Boltzmann-gas-theory,SEP-Boltzmann} for dilute gases.  It says that H function, which can be identified with $-S_{\rm T}$ for dilute gases in equilibrium, decreases monotonically, and achieves minimum at thermal equilbrium.  His kinetic equation was, however, criticized by Thomson and by Loschmidt~\cite{Ehrenfest,Boltzmann-biography,SEP-Boltzmann} for violation of time-reversal symmetry.   Burbury~\cite{Burbury-1894} and Bryan~\cite{Bryan-1894} later pointed out that the assumption of molecular chaos (or Stoßzahlansatz), which is used in deriving the collision term, breaks the time-reversal symmetry.  In response to these criticisms, Boltzmann argued that H theorem is valid only probabilistically~\cite{Ehrenfest,Boltzmann-biography,SEP-Boltzmann}: ``$\cdots$ if the initial state is not specially arranged for a certain purpose, but haphazard governs freely, the probability that H decreases is always greater than that it increases $\cdots$''.  This seems to imply a {\em stochastic version of H-theorem}, which, however, has never been established explicitly.  

%We note that Boltzmann's H function was defined on {\em individual coarse-grained microscopic states}.  

Boltzmann's H function is applicable only for dilute gases.  For general macroscopic systems, Boltzmann ``defined'' {\em a macro-state} as { a collection of micro-states that are {\em macroscopically indistinguishable}, and entropy via his famous formula~\cite{Boltzmann-biography}:
$S_{\rm B} = \log {\mathscr W}$, where $ {\mathscr W}$ is the number of these micro-states.   
%\label{Boltzmann-S-def}
%\vspace{-1mm} \ee 
$S_{\rm B}$ shall be called {\em Boltzmann entropy}.  Closely associated with this formula is Boltzmann's {\em fundamental postulate of equal a priori probability}, which says that all ${\mathscr W}$ states within the same collection are equally likely.  Note, however, Boltzmann's ``definitions'' are {\em not} a  definition in rigorous sense, unless the term ``macroscopically indistinguishable'' is made precise.   As Einstein remarked~\cite{Boltzmann-biography,Einstein-1909}: ``Neither Herr Boltzmann nor Herr Planck has given a definition of ${\mathscr W}$.''  Similar view was also expressed by R. Penrose \cite{Penrose}.  Very little is known about evolution of Boltzmann entropy in general~\footnote{Goldstein and Lebowitz~\cite{Lebowitz-2004} proved that Boltzmann entropy increases monotonically, if all relevant macroscopic quantities evolve deterministically and autonomously. We shall show however the premise does not hold: the macroscopic quantities generically stochastically. }. 
% \pagebreak.  
% Even though there have been many works following Boltzmann, 

In recent decades, studies of thermostated, dissipative non-equilibrium systems have led to a number of general  results called {\em Fluctuation Theorems} (FTs)~\cite{Bustamante-PhysToday-2005,Seifert-RPP-2012,Sevick-ARPC-2008,Gallavotti-FT,Lebowitz-FT-1999,Crooks-PRE-1999}, which relate the probability of a {\em process} $\Pi_+$ with {\em entropy production} $\Delta S$ to that of the time-reversed process $\Pi_-$ with entropy production $- \Delta S$.  The most general form of these theorems appears to be: 
\be 
{\rm  Pr} (\Pi_+) \, e^{-\Delta S} = {\rm Pr} (\Pi_-),  
\label{Crooks-FT-0}
\ee 
where $\Pi_{\pm}$ are macroscopic processes that start either from equilibrium, or from a steady state.  Hence entropy production may be either positive or negative.  Similar results were actually obtained by Bochkov and Kuzovlev as early as 1970's and 1980's~\cite{BK-1977}.  It is however often not clear how the entropy production $\Delta S$ is related to the change of total entropy.  To obtain a rigorous resolution of this issue, Liouville theorem can not be evaded.

%None of these works tell us what entropy is for an isolated system out of equilibrium. 

% None of these works however seem resolves the conflict between Liouville theorem and the second law.  
% In deriving these results, either noise or dissipation, or both, has to be introduced by hand.  It is not yet clear whether these Fluctuation Theorems are applicable to closed Hamiltonian systems.  

% Neither do they tell us whether (and why) entropy of a closed system increases monotonically with time.  

It may be attempting to interpret negative entropy production, which is allowed according to Eq.~(\ref{Crooks-FT-0}), as spontaneous violation of second law due to fluctuations.  However, careful inspections always indicate that decrease of entropy is  associated with measurements.  To apply Eq.~(\ref{Crooks-FT-0}), for example, one needs to  measure the initial and final states of the processes $\Pi_{\pm}$. The more refined the measurements, generically the lower entropy the resulting state.  Possibility of negative $\Delta S$ in Eq.~(\ref{Crooks-FT-0}) is therefore no more (and no less) than the paradox of Maxwell Demon, which can be resolved once the information acquired by the observer is taken into account~\footnote{If we consider the combined system including the system and the observer, the total entropy never decrease, even though correlations build up during the course of measurement.  If we only consider the system being studied, then the entropy should be defined as the conditional information entropy~\cite{Cover-Thomas-info-theory}, given all measurement results.  This entropy of course may decrease during the course of measurement.  But since the system is perturbed by the measurement, it is not closed, the second law is simply inapplicable.  }.  In general, observers have a substantial degree of control over entropy via measurements, a fact that may be called the {\em anthropomorphic principle of entropy}~\cite{Jaynes-1965-entropy}.    Hence before talking about the proper definition and evolution of entropy, one must specify how the system is being measured.  

%Neither Gibbs' idea of coarse-grained entropy nor Boltzmann's ``definition'' of entropy was formulated precisely.  This made it difficult to quantitatively analyze the evolution of their entropies, or to study the connections and differences between these concepts.  
In summary, our understanding of the second law and macroscopic irreversibility is still limited to heuristic reasoning and qualitative arguments, which more or less resemble that of Boltzmann in his late years~\footnote{For several very enlightening discussions on this issue, we refer the readers to the review papers by Lebowitz~\cite{Lebowitz-1993}, the books by Zwanzig ~\cite{Zwanzig-book},  by Penrose~\cite{Penrose}, and by Halliwell, Perez-Mercader, and Zurek~\cite{book-origin-time-asym}.  }.  At this stage, we lack evidence for decrease of entropy, just as we lack proof for monotonic increase of entropy.  We do not yet know whether macroscopic irreversibility is a reality, or merely an illusion due to our short life time.  To make progress, we will first give a precise reformulation of the seminal ideas by Boltzmann and by Gibbs, and clarify the proper definition of statistical entropy in different non-equilibrium experimental situations.  

\section{Unification of Boltzmann and Gibbs}

{\bf Boltzmann's Macro-states and Entropy}\quad  We shall focus on classical Hamiltonian systems exclusively in this work.  Discussion of quantum systems will be reserved for a separate publication. 
We use $\xv = (\qv^N, \pv^N)$ to denote a  micro-state, the totality of which forms the {\em phase space} $\Omega$.  Here $\qv^N, \pv^N$ are canonical coordinates and momenta. Following the notations by Goldstein and Lebowitz~\cite{Lebowitz-2004}, we introduce a {\em finite} set of mechanical quantities $\{ {\mathscr M}(\xv)\} = \{{\mathscr M}_1(\xv), \ldots, {\mathscr M}_m(\xv)\}$, which shall be called  {\em the macroscopic quantities}, to partition the phase space into cells, which shall be labeled by capital Roman letters, $A, B, C$, or $A_i$, etc, and shall be called {\em Boltzmann cells}.   Neighboring cells differ in macroscopic quantities $\{ {\mathscr M}(\xv)\}$ by at most  an amount $\epsilon$, which is in principle set by the resolution limit of measuring apparatus.  Every micro-state $\xv$ belongs to one and only one cell $A$, while the union of all cells is the entire phase space:
%\be
${\cup_i} A_i = \cup_A  A =  \Omega, 
\,\, A_i \cap A_j = A_i \, \delta_{ij},$
%\nonumber\\
%\label{A-partition}
%\ee
where $\cup, \cap$ denote union and intersection of sets.  We shall choose ${\mathscr M}_1 = E(\xv) = H(\qv^N,\pv^N)$, the intrinsic Hamiltonian that excludes the external forces.  We shall also assume that all other ${\mathscr M}_k$s are non-conserved~\footnote{If there are other conserved quantities, the phase space volume can be decomposed into multiple ergodic components.  The results derived in this work can be applied to each component.}.   The phase space can be decomposed into a sequence of shells $\Omega(E)$ with approximately constant energy:
%\begin{subequations}
%\ba
$\Omega(E) = \underset {\textit {\tiny A, E(A) =E}} {\cup} A, 
\,\, 
\Omega = \underset {\textit {\tiny E}} {\cup}  \, \Omega(E)$. 
Each of these shells form an ergodic component if the system is the Hamiltonian is time-independent.  Partition of phase space~\footnote{ A classical example of this construction is given for dilute gas by Boltzmann, where every cell corresponds to a coase-grained f-distribution $f(\vec{r},\vec{v})$, with $\vec{r}, \vec{v}$ position and velocity of single particle.   We note that total number of Boltzmann cells is generally countably infinite. }   is schematically illustrated in Figure \ref{Fig:partition-1.pdf}.

%Our result, Theorem \ref{thm-1}, is however valid independent of this issue. 
 
Following Boltzmann, all micro-states within a single cell $A$ will be {\em defined as} macroscopically indistinguishable, the totality of which forms a Boltzmann macro-state.   We assign a probability distribution to all these states, according to  {\em the fundamental postulate of equal a priori probability}:
\ba
\rho_A(\xv)  &=& \frac{1}{|A|} \chi_A(\xv) 
 \equiv \left\{
\begin{array}{ll}
 {1}/{|A|} , \quad&  \xv \in A,
\vspace{1mm} \\
0, & {\xv \notin A}.  
\end{array}
\right. \label{rho_A-def}
\ea
where |A| is the {\em dimensionless} volume of cell $A$ (see Eq.~(\ref{Boltz-Entropy-def}) for an explanation) and $\chi_A(\xv)$ is known as the {\em characteristic function} of set $A$.  The significance of Eq.~(\ref{rho_A-def}) is that the system achieves {\em conditional equilibrium} within the Boltzmann cell $A$.  Below we shall use $\rho_A(\xv)$ and $A$ interchangeably to denote a Boltzmann macro-state.   The entropy of $\rho_A(\xv)$  is already defined above.  For Hamiltonian systems,  ${\mathscr W}$ must be replaced by the dimensionless phase space volume:
\be
S_{\rm B}(A) = \log |A| \equiv \log {{\rm Vol}(A)}/{h^{d N}},
\label{Boltz-Entropy-def}
\ee
where $h^{dN}$ is a  microscopic  unit of phase space volume.  Correspondingly, integration over phase space will also be made dimensionless below.  
% It is also important to use canonical variables $\xv = (\qv^N,\pv^N)$ so that the integration measure in invariant under canonical transformations.  

\begin{figure}[t!]
\begin{center}
%\vspace{-5mm}
%\subfigure[]{	%}
		\includegraphics[width= 7.5cm]{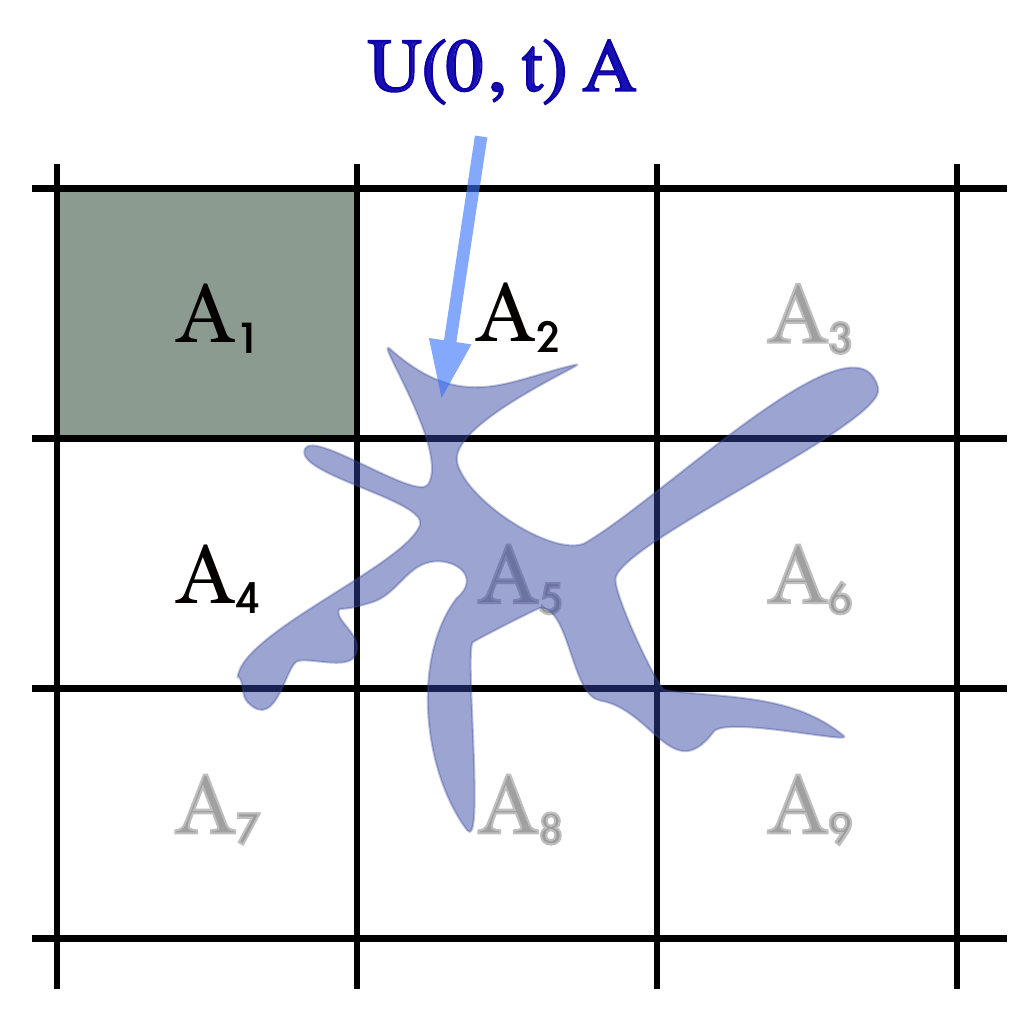}
\caption{Schematics: partition of phase space using Boltzmann cells.     The blue complex shape denotes an evolved set $U(0,t) A$, which overlaps with many cells.  The corresponding phase space pdf is given by Eq.~(\ref{rho-t-1-0}).  The volume $|U(0,t)A |$ is conserved during evolution, according to the Liouville theorem.  We have adjusted the scale such that all cells {\em appear of equal size. }  In reality, their volumes differ drastically. }
% In the limit $t \rightarrow \infty$, the set $U(0,t) A$ becomes a {\em fractal}.  }
\label{Fig:partition-1.pdf}
\end{center}
\vspace{-5mm}
\end{figure}

%The Planck constant $h$ has no significance in this work other than making $|A|$ dimensionless. 
 \vspace{2mm}

  {\bf Gibbs' Coarse-graining} \quad A {\em fine-grained distribution} is defined as a properly normalizednon-negative function $\rho(\xv)$ in $\Omega$. The {\em Gibbs entropy} of such a distribution  is already defined in Eq.~(\ref{S-def-Gibbs}).  A Boltzmann macro-state Eq.~(\ref{rho_A-def}) can be understood as a Gibbs state with pdf given by Eq.~(\ref{rho_A-def}), and its Gibbs entropy coincides with its Boltzmann entropy Eq.~(\ref{Boltz-Entropy-def}).   Following Gibbs's original idea, a fine-grained distribution $\rho(\xv)$ can be {\em coarse-granied} by performing local average over each cell $A$.  The resulting {\em coarse-grained} distribution $ \tilde{\rho}({\xv})$ is: 
\vspace{-2mm}
\begin{subequations}
\label{CGS-def}
\ba
&& \tilde{\rho}({\xv})
\equiv {\mathscr C} \rho(\xv) 
= \sum_A p_{\rho} (A) \rho_A(\xv),
\label{Pr(A)lpha-def} \\
%= (\chi_A, \rho), \\
%&& \sum_A Pr(A) = 1.
%\label{Pr(A)lpha-def-1} 
&& p_{\rho}(A) = \int_A  \rho(\xv) \,d \omega
= \int_A  \tilde{\rho}(\xv) \,d \omega
 =p_{\tilde{\rho}}(A) ,
\label{Pr(A)lpha-def-2} 
%&& \sum_A p(A) =1 
%\leftrightarrow \int \rho(\xv)d \omega = 1,
%\nonumber
\ea
\end{subequations}
where $p_{\rho}(A)$, being normalized as $\sum_A p_{\rho}(A) =1$, is the probability that the system is in Boltzmann state $A$, and ${\mathscr C}$ is the {\em coarse-graining operator}, whose mathematical definition is given in App.~\ref{app:Math}. 

%Two Gibbs states are said to be {\em macroscopically equivalent} if their coarse-grained version is identical.  It is in this sense that Boltzmann's fundamental postulate of equal probability remains applicable during the course of evolution, even if the fine-grained state $\rho(t)$ is non-uniform inside every Boltzmann cell.  

% We shall call two Gibbs states {\em macroscopically equivalent} if they yield identical averages for all chosen macroscopic quantities $\{ {\mathscr M}(\xv)\}$, or if they become identical after coarse-graining.  Some mathematical details are 

Using Eq.~(\ref{Pr(A)lpha-def}), the Gibbs entropy of a coarse-grained distribution $\tilde{\rho}$ can be written as:
\ba
S_{\rm G}[\tilde{\rho}] &=& 
-  \int_{\Omega} \tilde{\rho}({\xv}) \log  \tilde{\rho}({\xv}) d \omega 
\nonumber \\
&=&   \sum_{B} p_{\rho}(B) \log {|B|}
- \sum_{B} p_{\rho}(B) \log {p_{\rho}(B)}. 
 \label{S_G-C-G}
 \ea
Further using Eq.~(\ref{Pr(A)lpha-def-2}) we can show that $S_{\rm G}[\tilde{\rho}]$ is always larger than the fine-grained Gibbs entropy $S_{\rm G}[{\rho}]$:
\ba
\hspace{-5mm}
S_{\rm G}[\tilde{\rho}]  - S_{\rm G}[{\rho}]
%&=& -\sum_A \int_A d \omega  \left(
%\tilde{\rho} \log \tilde{\rho} 
%- \rho \log \rho\right)\nonumber\\
%&=&   - \sum_A \int_A d \omega  
%\left( {\rho} \log \tilde{\rho} 
%- \rho \log \rho \right)
% \nonumber\\
&=&   \int d \omega 
\, \rho \log \left( \frac{\rho} {\tilde{\rho}} \right)
\equiv D(\rho || \tilde{\rho})
\geq 0,
\label{S_G-C-G-2}
\ea
where $D( {\rho} || \tilde{\rho} ) $ is the {\em relative entropy} of $ \tilde{\rho}$ with respect to $ {\rho}$, which is known to be non-negative~\cite{Cover-Thomas-info-theory}. Hence coarse-graining always increases entropy. 
 
The macroscopic properties of a fine-grained distribution ${\rho}$ are completely encoded in its coarse-grained version $\tilde{\rho}$, the latter shall be called a {\em macro-state}. \footnote{Mathematically, we see that a macro-state corresponds an equivalent class of phase space pdf.  All distributions inside the same class become identical after coarse-graining. }  Accordingly, a system is said in equilibrium if its coarse-grained pdf converges to the equilibrium Gibbs distribution, or equivalently,  all macroscopic physical quantities obey equilibrium statistics.  Furthermore, Eq.~(\ref{Pr(A)lpha-def}) shows that any macro-state can be written as a linear superposition of Boltzmann states.  Therefore Boltzmann macro-states are building blocks of macro-states.   The operation of coarse-graining can  be understood as a special case of {\em renormalization transformation} which average out short scale details, but leave long scale properties intact.  

\vspace{2mm}
{\bf Anthropomorphic principle revisited}\quad We have defined three types of entropy: Boltzmann entropy, fine-grained Gibbs entropy, as well as coarse-grained entropy.  Which  one corresponds to the thermodynamic entropy for macroscopic systems out of equilibrium? {\em According to the anthropomorphic principle of entropy, the proper choice depends on how the system is measured.}  If the system is in certain Boltzmann macro-state $A$ (as a consequence of appropriate measurements), then Boltzmann entropy Eq.~(\ref{Boltz-Entropy-def}) is clearly the relevant concept.  If we do not know precisely the Boltzmann state, then the coarse-grained Gibbs entropy Eq.~(\ref{S_G-C-G}) should be used.  It is never correct to use the fine-grained Gibbs entropy Eq.~(\ref{S-def-Gibbs}), unless it agrees with the coarse-grained entropy.  

We have precisely formulated the seminal ideas of Boltzmann and of Gibbs using phase space partition, which apparently depends on {\em our choice} of macroscopic variables.   A moment of careful thinking indicates that by making this choice, we are hand-drawing a borderline between microscopic and macroscopic.  The fundamental justification of this choice comes from consideration of space/time scales.  We as creatures of macroscopic size (or our measuring apparatus) are not able to perceive spatial/temporal structures at microscopic scales.   A proper choice of $\{ {\mathscr M }_k \}$  should include all independent  extensive variables that do not achieve equilibriums in the time windows we can probe, together with possible localized variables that exhibit slow dynamics.   This is another reflection of the anthropomorphic principle.  
%This limitation ultimately justifies the construction of Boltzmann macro-state, as well as Gibbs' idea of coarse-graining.  \xing{The fundamental postulate and coarse-graining are two complementary views of the dichotomy between microscopic and macroscopic. }

%\vspace{1mm}

\section{Evolutions of Statistical Entropy}

{\bf Dynamics and Time-reversal} \quad  We shall be slightly more ambitious than Gibbs and Boltzmann, and study Hamiltonian systems driven by external forces which are {\em not} necessarily conservative.  We shall however assume that the external forces are independent of momenta, so that Liouville theorem is always valid.  For the formalism of Hamiltonian dynamics, see App.~\ref{app:work}.  

 We define an evolution operator $U(t_1, t_2)$, such that $U(t_1, t_2) \xv $ is the evolved micro-state at time $t_2$, if $\xv$ is the initial micro-state $\xv$ at $t_1$.    $U(t_1, t_2)$  can also be acted on a set $A$, or on a probability distribution $\rho(\xv)$:
\begin{subequations}
\ba
U(t_1, t_2)A &\equiv& \{U(t_1, t_2) \xv | \xv \in A\}, \\
U(t_1, t_2) \rho(\xv) &\equiv& \rho(U(t_1, t_2) ^{-1} \xv). 
\ea
\label{U-A}
\end{subequations}
Because of the chaotic nature of Hamiltonian dynamics, the evolved set $UA$ and the evolved function $\rho(U^{-1} \xv)$ become exceedingly more complicated as $t_2$ increases.  Nonetheless, Liouville theorem says that phase space volume and Gibbs entropy are invariant under evolution: 
%\begin{subequations}
 \ba
 |U(t_1, t_2) A | &=& |A|,
%  \label{Liouville-thm-1} \\
\quad S_{\rm G}[U(t_1,t_2)\rho] =  S_{\rm G}[\rho]. 
%  \label{Liouville-thm-3}
 \label{Liouville-thm}
 \ea
%\end{subequations}

We shall also define the time reversal of of a micro-state $\xv \equiv (\qv^N, \pv^N)$ as $\xv^* \equiv (\qv^N, - \pv^N)$.  The time reversal of a set $A$ and function $\rho(\xv)$ are defined  as $A^* \equiv \{ \xv^* | \xv \in A \}$ and $\rho^*(\xv) \equiv \rho(\xv^*)$.    Phase space volume and Gibbs entropy are also invariant under time-reversal:
\be
|A| = |A^*|, \quad 
 S_{\rm G}[\rho] =  S_{\rm G}[\rho^*]. 
 \label{T-reversal-inv}
\ee
We shall choose all macroscopic quantities  $\{ {\mathscr M }_k \}$ to be either even or odd order time-reversal.  Consequently, time reversal of a Boltzmann cell $A^*$ is another Boltzmann cell in the partition~\footnote{Of course, $A^*$ may or may not be the same cell as $A$. }, and coarse-graining commutes with time reversal: 
\be
{\mathscr C} \rho^* =({\mathscr C} \rho)^*.  
\label{C-T}
\ee
These identities will be useful for the derivation of  {\em stochastic H-theorem}. 
 
%\vspace{1mm}

{\bf Evolution of Boltzmann Entropy} \quad With a complete mathematical formalism in hand, we shall now study the evolution of statistical entropy.  Consider a typical experimental scenario, where we first fix a set of physical quantities $\{ {\mathscr M }_k \}$  and let the system equilibrate.  The initial state is then a Boltzmann state $\rho_A(\xv)$,  Eq.~(\ref{rho_A-def}).  We shall then relax certain constraint, or apply some external forces, and let the system evolve from $t= 0$.  The fine-grained distribution  at time $t$ is
\begin{subequations}
\ba
\rho(\xv,t) = U(0,t) \rho_A(\xv) &=& \rho_A(U^{-1} \xv). 
\label{rho-t-1-0}
\ea
Its coarse-grained version  can be written as  
\be
\tilde{\rho}(\xv, t) =  
{\mathscr C }U(0,t) \rho_A(\xv)
\equiv \sum_B{Pr}(B, t |A, 0) \, \rho_B(\xv), 
\label{tilde-rho-1-0}
\ee
\end{subequations}
where ${Pr}(B, t |A, 0)$, as illustrated in Fig.~\ref{Fig:partition-1.pdf}, is  the volume fraction of  $ U(0,t)  A$ intersecting $B$: 
 \ba
{Pr}(B, t |A, 0) %&=& \int_{B} d \omega \,  \rho(\xv, t)
%=  \frac{1}{|A|} (\chi_B, U \chi_A)
%\nonumber\\ 
=   |U(0,t) A \cap B|/{|A|}. 
\label{PAB}
\ea 
For a detailed derivation of this result, see Appendix \ref{app:PAB}.   

There are generically many terms in the RHS of Eq.~(\ref{tilde-rho-1-0}), which means that {\em Boltzmann states evolve stochastically}.  If we measure all macroscopic quantities again at time $t$, and determine the Boltzmann state.   According to Eq.~(\ref{tilde-rho-1-0}), the probability of obtaining Boltzmann state $B$ is ${Pr}(B, t |A, 0)$, which called the {\em transition probability} from Boltzmann states $A$ to  $B$.  

We shall now consider the {\em backward dynamics}, with both the Hamiltonian and the external forces time-reversed.  Let the backward evolution operator be $U^*(t_1,t_2)$. The system evolves from a state $\xv$ to another state $\yv$ under forward dynamics, if and only if it evolves from ${\yv^*}$ to $\xv^*$ under the backward dynamics.  We shall now consider the {\em backward macroscopic dynamic process} where the system starts from Boltzmann state $B^*$ and transits to state $A^*$. The corresponding transition probability ${{Pr}^*}(A^*,t|B^*; 0)$ can be obtained from Eq.~(\ref{PAB}) by simple replacements  $(A, B, U(0,t)) \rightarrow (B^*, A^*, U^*(0,t))$:
 \ba
{{Pr}^*}(A^*, t|B^*,0)
&=& {| U^*(0,t)B^* 
\cap  A^*|}/{|B^*|}. 
\label{PBA}
\ea
Using the invariance of Liouville measure under time reversal and evolution, we can obtain a simple relation between the probabilities of the forward and backward processes (Details of proof are given in Appendix \ref{app:PAB}):
\begin{theorem}
\label{thm-1} 
( Stochastic H-Theorem)
 For closed Hamiltonian systems, evolution of Boltzmann entropy is stochastic and satisfies the following relation:
\ba
{Pr}(B, t|A, 0 ) &\equiv& e^{S_{\rm B}(B) - S_{\rm B}(A)}
{{Pr}^*}(A^*, t|B^*,0),
 \label{second-law}
\ea
%where $A, B$ are, respectively, the initial and final Boltzmann states of the forward transition. 
\end{theorem}
where $S_{\rm B}(A) = \log |A|, S_{\rm B}(B) = \log |B|$ are Boltzmann entropies.  The term {\em Stochastic H-Theorem} is mort appropriate, because Eq.~(\ref{second-law}) rigorously quantifies Boltzmann's late understanding that evolution of H-function is stochastic, and generalize it to arbitrary interacting Hamiltonian systems.  We stress again that the possibility of $S_{\rm B}(B) < S_{\rm B}(A) $ is a consequence of measurement, and should not be understood as failure of the second law.  

%\xing{In fact, using Theorem \ref{thm-1}, we can easily prove Planck's statement of second law (Planck’s Principle): The internal energy of a closed system is increased by an adiabatic process (Volume fixed). Here adiabatic means there is no exchange of heat between the system and the environment. }

The Stochastic H-Theorem can be understood as a general version of Fluctuation Theorems for closed Hamiltonian systems.  Since Boltzmann states are building blocks of generic macro-states, and heat bath can always be included to form a closed system, Eq.~(\ref{second-law}) can be used to construct various Fluctuation Theorems and related identities.  This is done in App.~\ref{app:FTs}.  It is important to note that study of Fluctuation Theorems and their implications is a vast subject, and it is not our purpose to undermine their importance or priority.  Rather, by establishing their connections with Boltzmann's work through Theorem \ref{thm-1}, we enlarge their domain of applicability, as well as provide for them a more reliable foundation.   
% We shall be satisfied with a brief discussion of their relation with Theorem \ref{thm-1}.  

%In App.~\ref{app:Steady-State}, \ref{app:Crooks}, \ref{app:Evans-Searles}, and \ref{app:BKE-JE}, we shall further use it to derive various Fluctuation Theorems and work identities developed in the recent years.  

% Later, we shall explicitly derive from Eq.~(\ref{second-law}) various identities by Bochkov and Kuzovlev, by Evans and Searles, and also by Crooks. 

%The powerfulness of Eq.~(\ref{second-law}) originates from the flexibility and generality of Boltzmann's notions of macro-state and entropy. 

% It can be understood as a unification of various Fluctuation Theorems and Work Relations derived in recent decades.
 
%Let us first consider an isolated system in equilibrium. The dynamics is governed by a time-independent Hamiltonian, hence $\tilde{{ P}} = { P}$.  We choose all $\{ {\mathscr M}_k \}$ to be even under time-reversal, so that $A^* =A, B^* = B$.  Then Eq.~(\ref{second-law}) reduces to well-known condition of detailed balance:
%\be p_{\rm EQ}(A) {Pr}(B, t |A, 0 )
%= p_{\rm EQ}(B){Pr}(A, t|B, 0 ), \ee
%where $p_{\rm EQ}(A) = |A|/|\Omega|$ is the equilibrium probability of macro-state $A$.  %This is, of course, the well known relation of detailed balance. 

%\label{sec:Second-Law} 

\vspace{1mm}

{\bf Evolution of Coarse-grained Gibbs Entropy} \quad  Now let us consider a different experimental scenario, where the system starts from a Boltzmann state $\rho_A$, and keep evolving without being measured.  The relevant entropy is then the coarse-grained Gibbs entropy Eq.~(\ref{S_G-C-G}), with $\tilde{\rho}$ given by Eq.~(\ref{tilde-rho-1-0}). The question is whether the coarse-grained Gibbs entropy increases monotonically with time.

First of all, since the initial distribution $\rho_A(\xv)$ is invariant under coarse-graining, the following result is easy to establish and was in fact already known by Gibbs~\cite{Gibbs-principle}\footnote{Note, however, Ehrenfest believed that Gibbs' exposition was incorrect.  See Sec. 27, page 71 of reference \cite{Ehrenfest}. }: 
\be
S_{\rm G}[\tilde{\rho}(t)]  \geq
 S_{\rm G}[{\rho}(t)] = S_{\rm G}[{\rho}(0)]
 = S_{\rm G}[{\tilde \rho}(0)], \,\, 
\forall \,\, t > 0.
\label{Gibbs-inequality}
\ee
 Note that we have used successively Eqs.~(\ref{S_G-C-G-2}) and (\ref{Liouville-thm}).  This result is however not the second law, for that would require that $S_{\rm G}[\tilde{\rho}(t)]$ monotonically for all $t$.  

%This is the central problem we will study in this section.  
% In this section, we shall prove this result, under two assumptions: (1) the Hamiltonian dynamics satisfies certain {\em mixing} proprty, and (2) $|t' - t| \gg \tau_{\rm mixing}$ where $\tau_{\rm mixing}$ is the microscopic mixing time scale.   

%\subsection{Ensemble of Coarse-grained Trajectories, and Mixing}
%\label{sec:formalism-KS}
%The coarse-grained state $\tilde{\rho}(\xv,t)$ and the fine-grained state $\rho(\xv,t)$ are equivalent at the macroscopic level, in the sense that  they yield the same averages for all chosen observable quantities $\{ {\mathscr M}_k\}$.

Since all we care are macroscopic properties, it would be convenient if we can characterize the evolution of $\tilde{\rho}(t)$ without referring to ${\rho}(t)$.  Unfortunately this is impossible in general, because coarse-graining usually leads to loss of information.  If however the following operator identity is valid
\begin{subequations}
\be
\lim_{t,\tau \rightarrow \infty}
{\mathscr C}  U(0, t + \tau) {\mathscr C}
 = \lim_{t ,\tau\rightarrow \infty} {\mathscr C}
  U(\tau, t + \tau)  {\mathscr C} U(0,  \tau){\mathscr C} , 
\label{local-mixing-def-0}
\ee
we would have the desired property in the limit $t,\tau \rightarrow \infty$, as long as the system starts from a Boltzmann state, or a linear combination thereof.  Let Eq.~(\ref{local-mixing-def-0}) acting on $\rho_A$, we obtain
\ba
\lim_{t,\tau \rightarrow \infty} \tilde{\rho}(t + \tau) 
%&=&  {\mathscr C}  {\rho}(t + \tau) 
%=  {\mathscr C} U(\tau, t +\tau)  \rho(\tau)
%\nonumber\\
%= {\mathscr C} U(\tau, t +\tau) {\mathscr C} \rho(\tau)
 =\lim_{t,\tau \rightarrow \infty} {\mathscr C} U(\tau, t +\tau) \tilde{\rho}(\tau),
\,\, \label{local-mixing-def-1}
\ea
\label{local-mixing-def}
\end{subequations}
where ${\mathscr C} U(\tau, t +\tau)$ is the evolution operator for coarse-grained distribution. 

 We say that the microscopic dynamics is {\em locally mixing}, if the operator identity Eq.~(\ref{local-mixing-def-0}) hold.  Qualitatively speaking, local mixing means that microscopic details (the differences between $\rho(\tau)$ and $\tilde{\rho}(\tau)$) at time $\tau$ do not influence macroscopic properties at any remote future time $t + \tau$, as long as the state at $\tau$ is itself a consequence of long evolution from a Boltzmann state.  In App.~\ref{app:mixing} we discuss the  differences and connections between local mixing and the mixing property frequently studied in ergodic theory~\cite{book-Sinai-Ergodic theory}.   {\em The most important point here is that if a system approaches equilibrium, it must be locally mixing.}  Since most realistic Hamiltonian systems we know indeed equilibrate, we are not really losing anything by assuming the system to be locally mixing.  We also note that the characteristic time scale $\tau_{\rm LM}$ of local mixing (when the limits in Eq.~(\ref{local-mixing-def-0}) converge) should be mesoscopic (i.e., remains finite as the system becomes large), if all slow variables are already included in the list of macroscopic quantities $\{{\mathscr M}_k\}$.  Combining Eqs.~(\ref{local-mixing-def-1}), (\ref{S_G-C-G-2}), and (\ref{Liouville-thm}), we see that for $t ,\tau \gg \tau_{\rm LM}$:
\ba
S_{\rm G}[\tilde{\rho} (t + \tau) ] 
&=& S_{\rm G}[{\mathscr C} U(\tau, t + \tau) \tilde{\rho}(\tau) ]
\nonumber\\
&\geq& S_{\rm G}[U(\tau, t + \tau) \tilde{\rho}(\tau)]
= S_{\rm G}[ \tilde{\rho}(\tau) ].
\label{Delta-S-relative-0}
\ea
Hence we obtain a theorem specifying the sufficient conditions for coarse-grained entropy to increase: 
\begin{theorem}
\label{thm-2}
(Second Law of Thermodynamics)  If the Hamiltonian dynamics is locally mixing, and the system start from a Boltzmann state, the coarse-grained Gibbs entropy $S_{\rm G}[\tilde{\rho}(t) ]$ increases monotonically in time scales much longer than the local mixing time.  
\end{theorem} 
Evolutions of Boltzmann entropy, coarse-grained Gibbs entropy and fine-grained Gibbs entropy for a locally mixing system are schematically illustrated in Fig.~\ref{Fig:entropy-evolution}.   Note that this theorem does not say anything about whether the system converges to an equilibrium state.  In fact, if the system is driven by external forces, it will never equilibrium.  Its average energy as well as its coarse-grained Gibbs entropy will keep increasing without bound.  On the other hand, if there is no driving force,  
it can be proven that a locally mixing system will eventually approach thermal equilibrium in each ergodic component of phase space (with some mathematical conditions, which are expected to be satisfied by usual physics systems).  In App.~\ref{app:Open systems}, we shall generalize Theorem \ref{thm-2} to open systems (in contact with a heat bath), and show that the non-equilibrium free energy decreases monotonically, as long as the combined system is locally mixing.  

%The proof will not be given, due to limitation of space. 

\begin{figure}[t!]
%\vspace{5mm}
\begin{center}
		\includegraphics[width = 9cm]{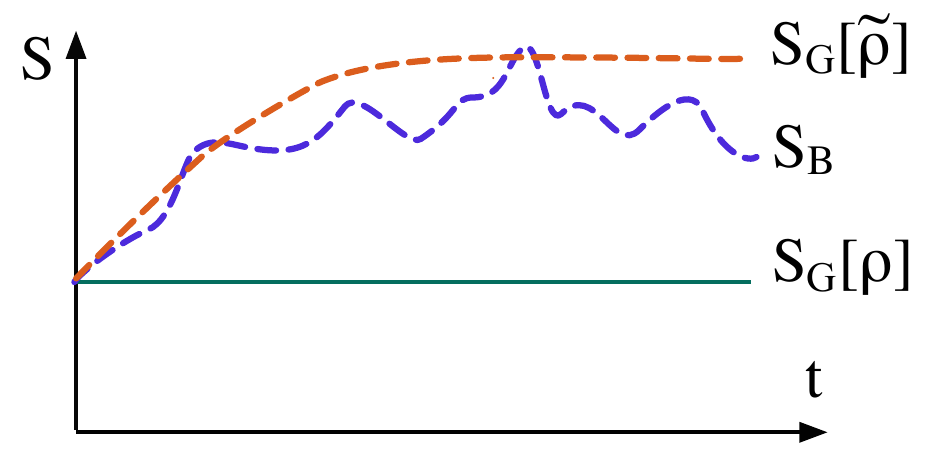}
\caption{From bottom to up, fine-grained Gibbs entropy is conserved; Boltzmann entropy fluctuates stochastically; coarse-grained Gibbs entropy (for a locally mixing system) increases monotonically. Note that evolution of Boltzmann entropy is defined as a discrete sequence of measurements, whilst monotonicity of coarse-grained Gibbs entropy is guaranteed only in  time scales longer than the local mixing time.  }
\label{Fig:entropy-evolution}
\end{center}
\vspace{-5mm}
\end{figure}

\vspace{2mm}
{\bf From Deterministic to Markov} \quad Equation (\ref{local-mixing-def-1}) says that $\tilde{\rho}(t + \tau)$ is fully determined by $\tilde{\rho}(\tau)$.  That is,  it does not depend on the more ancient history , as long as  $\tilde{\rho}(\tau)$ is given.  This is precisely the Markov property of stochastic processes.  If we discretize the time as $t_k = k \tau$ with $\tau \gg  \tau_{\rm LM} $, and expand $\tilde{\rho}_k \equiv {\rho}( t_k)$ in terms of Boltzmann states: 
\ba
\tilde{\rho}_k \equiv {\rho}( t_k)
&=& \sum_A p_k(A) \rho_A, 
\ea
and substitute this back to Eq.~(\ref{local-mixing-def-1}), we obtain the following recursive relation for  $p_k(A)$:
\be
p_{k+1}(A) = \sum_{B} {Pr}(A, (k+1)\tau | B, k \tau) p_k(B),
 \label{CK-2}
\ee
where ${Pr}(A, (k+1)\tau | B, k \tau)$ is the transition probability defined in Eq.~(\ref{PAB}).  In fact, one can easily prove that Eq.~(\ref{CK-2}) is equivalent to Eqs.~(\ref{local-mixing-def}), and hence can be used  as an alternative definition of local mixing.  But Eq.~(\ref{CK-2}) is the {\em Chapman–Kolmogorov} equation, which implies that the stochastic sequence of Boltzmann states $\{A(k\tau), k = 0, 1, 2, \ldots \}$ is a {\em Markov chain}.  Hence we have the following theorem: 
\begin{theorem}
\label{thm-3}
(Markov Chain of Boltzmann States) The  sequence of Boltzmann states $\{A(k\tau), k = 0, 1, 2, \ldots \}$ forms a Markov chain, if and only if the Hamiltonian dynamics is locally mixing, and $\tau \gg \tau_{\rm LM}$. 
\end{theorem}
This Theorem explicitly demonstrates how macroscopic stochastic evolutions can be fully consistent with microscopic deterministic evolutions.  It therefore provides a straightforward way of testing whether a particular manybody system is locally mixing, and whether the second law (Thm \ref{thm-2}) is applicable.  For an illustration, see Fig.~\ref{Fig:evolution-two-views}.  

\begin{figure}[tp!]
\begin{center}
		\includegraphics[width=10cm]{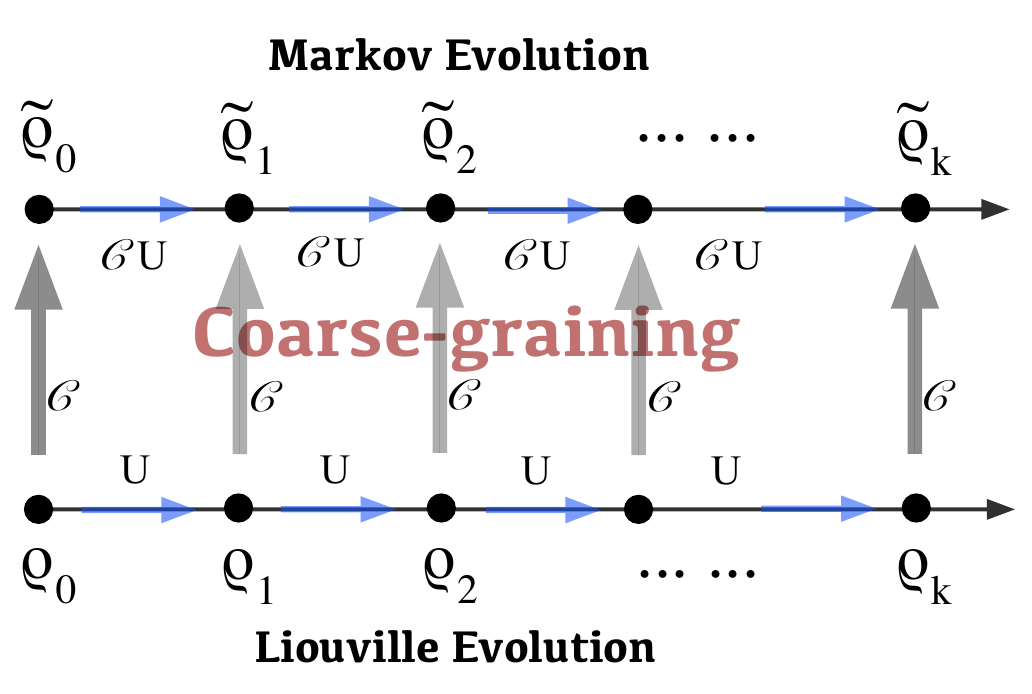}
\caption{For a locally mixing Hamiltonian system, the microscopic evolution is unitary and ruled by  Liouville equation, whereas the macroscopic evolution is Markovian and ruled by Chapmann-Kolmogorov equation.  Vertical arrows: coarse-graining.  }
\label{Fig:evolution-two-views}
\end{center}
\vspace{-5mm}
\end{figure}

We have already argued that the local mixing time $\tau_{\rm LM}$ is independent of system size. There is another time scale $\tau_{\rm EQ}$, the equilibrium time for the whole system, which diverges with as the system size becomes large.   For $t \ll \tau_{\rm LM}$, microscopic details prevail, and entropy does not necessarily increase, whereas for $t \gg \tau_{\rm EQ}$, the system equilibrates, and entropy no longer changes.  In the time window $\tau_{\rm LM} \ll t \ll \tau_{\rm EQ}$, the coarse-grained Gibbs entropy keeps increasing, and therefore the macroscopic physics is irreversible.   

%Hence we see that $\tau_{\rm LM}$ is the time scale of macroscopic irreversibility.  

%\vspace{2mm}
{\bf  Arrow of time} \quad Theorem \ref{thm-2} is manifestly asymmetric in time-reversal, and hence {\em gives  an arrow to the time} for classical statistical physics.  We note that the very definition of local mixing, Eq.~(\ref{local-mixing-def-0}), is symmetric under time-reversal, which means that if a dynamics is locally mixing, the time-reversed dynamics is also locally mixing.  This is in strong contrast with the assumption of {\em Molecular Chaos}, which breaks the time-reversal symmetry, as was pointed out by Burbury~\cite{Burbury-1894} and Bryan~\cite{Bryan-1894}.   So what creates the arrow of time in Theorem \ref{thm-2}? 
 
Let us explicitly construct a process where the coarse-grained Gibbs entropy {\em decreases} over time.  For simplicity, we consider an isolated system with time independent Hamiltonian, so the evolution operators for the backward and forward dynamics are identical, i.e., $U^*(t) = U(t)$.  Let the system evolve from $\rho_A$ and  $U(t)\rho_A$.  The final entropy is larger than the initial entropy, $ S_{\rm G}[{\mathscr C} U \rho_A] \geq S_{\rm G}[ \rho_A] $.   Now we consider the time-reversed process, where the system starts from the state $(U\rho_A)^*$ and evolves to $\rho_A^*$.  Mathematically, we have $U (U\rho_A)^* = \rho_A^*$.  The entropy of the new final state is indeed {\em smaller than} that of the new initial state, since using Eqs.~(\ref{Liouville-thm},\ref{T-reversal-inv},\ref{C-T}) we easily find
\ba
S_{\rm G}[{\mathscr C}(U\rho_A)^*] 
&=&S_{\rm G}[ {\mathscr C}U\rho_A ] 
 %S_{\rm G}[({\mathscr C}U\rho_A)^*] = 
%= S_{\rm G}[({\mathscr C}U{\mathscr C}\rho_A)] 
%\nonumber\\
\geq S_{\rm G}[U  \rho_A]
%=  S_{\rm G}[{\mathscr C}\rho_A] 
=  S_{\rm G}[\rho_A^*]. 
\nonumber
\ea
As pointed out by Zwanzig~\cite{Zwanzig-book}, bizarre behaviors may happen if {\em the initial conditions are created maliciously}.  In this example, the time-reversed non-equilibrium state $(U\rho_A)^*$ is the malicious initial data.  It is not realizable in experiments, but can be realized readily in computer simulations~\cite{T-reverse-simulation} by inverting all velocities\footnote{In quantum spin systems, time-reversed state can be realized {\em effectively} by inverting the magnetic field~\cite{spin-echo-Hahn}. }.  We note, however, if we let the backward dynamic process keep evolving from {\em the new final state} $\rho_A^*$, the entropy will start to increase again, as predicted by Theorem \ref{thm-2}, since $\rho_A^*$ is itself a Boltzmann state.  We also note that the issue of Poincar\`e recurrence needs not to be considered, simply because its time scale is too long to be experimentally relevant.    

So entropy will decrease if the system start from $(U\rho_A)^*$.  But it will increase if starting from $U \rho_A$, as guaranteed by Theorem \ref{thm-2}.  What is the difference between these two {\em initial states}? As far as I can see, the only difference is that $U \rho_A$ is evolved from a Boltzmann state $\rho_A$, whereas $(U\rho_A)^*$ is not.  Hence it is the {\em initial condition} as a Boltzmann state that is  responsible for the arrow of time in classical thermodynamics.  After all, states such as  $(U\rho_A)^*$ can not be prepared using macroscopic operations.   In fact, as hinted by R. Penrose, I tend to believe that all arrows of time, including those in cosmology, in causality, and in consciousness are ultimately results of special initial conditions.   

Some readers may think that the time arrow demonstrated above contradicts Poincare recurrence, which is known to appear for every system with finite phase space.  To see that this is not the case, we emphasize that recurrence is known to be valid for ergodic systems only at the level of single system.  Recurrence at the level of ensemble is however a completely different thing.  It would require the evolved set $U(t)A$ to become $A$ for some very large $t$, which is certainly not possible for generic cases. 

{\bf Connection to  Quantum Systems} \quad  In 1929 von Neumann \cite{Neumann-1929} \footnote{Also see a commentary~\cite{Lebowitz-2010-1} by Goldstein {\it etl.al.}, and a companion article which improves Neumann's first theorem~\cite{Lebowitz-2010-2}. }
studied the statistical mechanics of isolated quantum systems, and established two important theorems: the quantum ergodic theorem and the quantum H-theorem.  In the first theorem, he proved that all macroscopic quantities are almost always close to their equilibrium expectation values, whereas in the second theorem, he proved that the {\em coarse-grained von Neumann entropy} is almost always close to the equilibrium entropy.  No result however has been established for the monotonic increase of coarse-grained entropy.  In fact,  such a result cannot be valid for isolated quantum systems with finite size, because these systems (either in pure state or in mixed state) are known to be quasi-periodic, and hence must exhibit Poincare recurrence.   A realistic hope would be to establish the monotonicity of von Neumann entropy for systems in contact with an infinite heat bath, which shall be explored in a future work.  We note, however, the theoretical formalism of von Neumann also rely heavily on the ideas of coarse-graining and of macroscopic quantities, very much like the present work~\footnote{Perhaps one should not be so surprised by this similarity, since Neumann likely also drew inspiration from Gibbs' idea of coarse-graining. }.   In recent years, there has been a resurgence of interests in the equilibration of isolated quantum systems~\cite{Srednicki-ETH-1,Srednicki-ETH-2,Biao-QMxing-1,Biao-QMxing-2}.  Many of the results obtained so far seem to provide concrete verifications of two theorems of von Neumann. % We shall study evolution of entropy for quantum systems in more detail in the near future.  

\section{Concluding Remarks} 
By synthesizing the seminal ideas of Boltzmann and of Gibbs, and by clarifying the {\em anthropomorphic principle} of entropy, I have given a proper definition for statistical entropy for non-equilibrium Hamiltonian systems, and have resolved the conceptual conflict between the second law and Liouville theorem. I  have found that while local mixing property is responsible for the emergence of Markov dynamics at macroscopic level, it is the initial condition of Boltzmann state that is ultimately responsible for the arrow of time in classical statistical mechanics.  These results are general and completely independent of the details of Hamiltonian.

The author acknowledges Yongshi Wu, Tony Leggett, Ping Ao, Haijun Zhou, Biao Wu, Fei Liu, Haitao Quan, Xiaosong Chen, Zhanchun Tu for stimulating discussions.   This work is supported by NSFC via grant \#11674217, as well as Shanghai Municipal Education Commission and Shanghai Education Development Foundation via ``Shu Guang'' project.

%\bibliography{/Users/xxing/research/reference-all}

\pagebreak

\appendix

\section{Partitions and Coarse-grainings}
\label{app:Math}
The {\em characteristic function} of any set $A$ is defined as:
\be
\chi_A(\xv) = \left\{
\begin{array}{cc}
1,& \xv \in  A;
\vspace{2mm}\\
0, & \xv \notin  A. 
\end{array}
\right.
\ee
Because all Boltzmann cells $\{A\}$ form a partition of the phase space $\Omega$, we have 
\be
\sum_A \chi_{A}(\xv) = 1.
\ee 
Integral of $\chi_A$ over the phase space is just the volume of $A$:
\be
\int_{\Omega} \chi_A(\xv) d \omega = |A|,
\ee
where $d \omega = d^{3N} \! p  \, d^{3N} \!  q/h^{3N}$ is the dimensionless Liouville measure of the phase space, which is itself invariant under Hamiltonian dynamics.   

The inner product of two real-valued functions $f(\xv), g(\xv)$ is defined as:
\ba
(f, g) \equiv \int_{\Omega} f(\xv) g(\xv) d \omega.
\label{inner-prod-def}  
\ea
The following identities regarding characteristic functions are easily proved:
\begin{subequations}
\ba
(\chi_A, \chi_B) &=& |A \cap B| = |A| \,\delta_{AB}, \\
(U \chi_A, \chi_B) &=& | U A \cap B|. 
\ea
\label{U-chi-chi}
\end{subequations}
The evolution operator $U(t_1,t_2)$ as an operator acting on the phase space has determinant one (Liouville theorem):
\be
\det \left( \frac{\partial (U \xv)}{\partial \xv} \right) = 1. 
\ee 
When understood as an operator on the function space, $U$ is {\em unitary}, i.e., it preserves the inner product:
\be
(f, g) = (Uf, Ug) = (U^{-1}f, U^{-1}g).
\ee
This can be easily proved using Eq.~(\ref{inner-prod-def}) via a coordinate transformation.  

The probability $p_{\rho}(A)$ defined in Eq.~(\ref{Pr(A)lpha-def-2}) can then be represented as inner-product $\chi_A$ and $\rho$:
\be
p_{\rho}(A) = \int_A  \rho(\xv) \,d \omega
 = (\chi_A, \rho). 
\ee
The coarse-graining operator is already defined in  Eq.~(\ref{Pr(A)lpha-def}).  It can also be expressed in terms of $\chi_A$: 
\ba
({\mathscr C} f) (\xv) &=&  \tilde{f} (\xv) 
 = \sum_A\chi_A(\xv)
  \frac{1}{|A|} (\chi_A, f) 
\nonumber\\
&\equiv&
 \int_{\Omega} {\mathscr C}
(\xv, \yv) f(\yv)  d\omega_{y}.
  \label{C-def-1}
\ea
The kernel function ${\mathscr C}(\xv, \yv)$ can be expressed as 
\ba
{\mathscr C}(\xv, \yv) \equiv 
\sum_A \frac{1}{|A|} \chi_A(\xv)  \chi_A(\yv). 
\ea
It is easy to see that ${\mathscr C}$ is {\em self-adjoint} and {\em idempotent}:
\ba
(f, {\mathscr C} g) = ({\mathscr C}  f, g), \quad
{\mathscr C} ^2 f = {\mathscr C}  f. 
\ea
Hence it is a {\em projection operator}.
Furthermore it is easy to see that all $\rho_A, \chi_A$ are invariant under coarse-graining, i.e., they are eigenfunctions of ${\mathscr C}$ with eigenvalue unity:
\ba
{\mathscr C} \rho_A = \rho_A, \quad
{\mathscr C} \chi_A = \chi_A. 
\label{g-C-f}
\ea

The expectation value of a macroscopic quantity ${\mathscr M}$ over a Gibbs pdf $\rho(\xv)$ can be computed:
\ba
\langle {\mathscr M} \rangle_{\rho}
& \equiv& (M, \rho) 
= \sum_A \int_A M(\xv) \rho(\xv) d \omega. 
\label{M-rho}
\ea
But according to our construction, inside each cell $A$, $M(\xv)$ varies only by an infinitesimal amount. Hence we can replace $M(\xv)$ by its value at any point inside the cell, or by its average $M(A)$ inside the cell, defined as 
\be
 M(A) \equiv \frac{1}{|A|} \int_A M(\yv) d\omega_{y}
= \frac{1}{|A|} (\chi_A, M). 
\label{M-cell-approx}
\ee
This allows us to simplify Eq.~(\ref{M-rho}):
\be
\langle {\mathscr M} \rangle_{\rho} 
\approx  \sum_A  M(A ) p_{\rho}(A) 
%\sum_A  \frac{1}{|A|} (M, \chi_A) (\chi_A, \rho)
= ( M, {\mathscr C} \rho)
= ( M, \tilde{\rho}). 
\ee
Hence, as far as values of macroscopic quantities are concerned, there is no difference between $\rho(\xv)$ and its coarse-grained version $\tilde{\rho}(\xv) = {\mathscr C}\rho(\xv)$.   This is why we define a macroscopic state as a coarse-grained Gibbs distribution.  We say that two Gibbs states $\rho_1(\xv), \rho_2(\xv)$ are macroscopically equivalent if their coarse-grained versions are identical.

\section{Hamiltonian Dynamics and Time-Reversal}
\label{app:work}

For Hamiltonian systems driven by external forces, there are multiple ways that work and energy can be defined.  We shall however always define {\em work} as the change of energy.   To simplify the notations, we use $\qv, \pv$, instead of $\qv^N, \pv^N$ for the N-body canonical coordinates and momenta.  Let $H(\qv,\pv)$ be the Hamiltonian {\em in the absence of external force},  and is independent of time.  Let ${\mathbf f}_{\rm ext}(\qv,t)$ be the external force, which is generically non-conservative.  We shall assume that the external force ${\mathbf f}_{\rm ext}(\qv,t)$ is independent of momenta $\pv$.   We shall refer to $H(\qv,\pv)$ as the {\em intrinsic Hamiltonian}.  The Hamiltonian equations are:
\begin{subequations}
\label{Hamiltonian-eqn}
\ba
\dot\qv &=& \frac{\partial H}{\partial \pv}, \\
\dot\pv &=& - \frac{\partial H}{\partial \qv} 
+ {\mathbf f}_{\rm ext}(\qv,t). 
\ea
\end{subequations} 
The {\em velocity field} in the phase space is defined as 
\be
{\mathbf v}(\xv) = (\dot{\qv}, \dot{\pv}).
\ee
The divergence of this velocity field is
\ba
\nabla \cdot {\mathbf v}(\xv)
&=& \nabla_{\qv} \cdot \dot{\qv}
+ \nabla_{\pv} \cdot \dot{\pv}
\nonumber\\
&=&  \nabla_{\pv} \cdot  {\mathbf f}_{\rm ext}(\qv,t)
=0. 
%\nabla_{\qv} \cdot \nabla_{\pv} H 
%- \nabla_{\pv} \cdot \nabla_{\qv} H 
\ea
Therefore, as long as the external force is independent of momenta,  the velocity field in phase space is divergenceless, and phase space volume is conserved by the dynamics.  This is Liouville theorem.  

\subsection{Definitions of energy and work}
One natural definition of energy is the value of intrinsic Hamiltonian: 
\be
{ E} = H(\qv,\pv).
\ee  
Correspondingly, the {\em exclusive work} $W$ is defined as:
 \ba
d W = {d { E}}  % &=& {d H}
=  \frac{\partial H}{\partial \pv} \cdot d {\pv}
+  \frac{\partial H}{\partial \qv} \cdot d{\qv}
%\nonumber\\
= {\mathbf f}_{\rm ext}(\qv,t)  \cdot  d{\qv}. 
\label{dW-def}
\ea   
Integrating along a segment of trajectory, we find
\be
{ W}  \equiv  \Delta { E}
= \Delta H
=  \int_{0}^t {\mathbf f}_{\rm ext}(\qv,t) \cdot \dot{\qv} \, d t. 
\ee 
%where the integration is from the initial time to the final time, and $\qv, \dot{\qv}$ are evaluated along the trajectory.  
%\xing{Discuss the relation between ${\mathcal W} $ and dissipation. }

%This view is applicable for arbitrary external force, which may not be conservative.  If a system is both driven and thermostated, there is a constant energy input from the external force and energy output to the heat bath.  Hence $H_0$ and ${\mathcal W}$ both remain constants. 

It is always possible (and possibly convenient) to decompose the external force $ {\bf f}_{\rm ext}( \qv,t )$ into two parts:
\be
{\bf f}_{\rm ext}( \qv,t ) = - \nabla_{\qv} V(\qv, t )
 + {\bf f}'_{\rm ext}( \qv,t ). 
\label{f-decomp}
\ee
where the first part is {\em conservative} with $V(\qv,t)$ the external potential energy $V(\qv,t)$, and the second part $ {\bf f}'_{\rm ext}( \qv,t )$ is {\em non-conservative}.  The decomposition is of course not unique.   We can also define a {\em total Hamiltonian} $ H'(\qv,\pv,t)$ as:
\ba
 H'(\qv,\pv,t) &=& H(\qv,\pv) + V(\qv, t).  
\ea
The Hamiltonian equations (\ref{Hamiltonian-eqn}) can be rewritten as
\begin{subequations}
\label{Hamiltonian-eqn-1}
\ba
\dot{\qv} &=& \frac{\partial H'}{\partial \pv}, \quad 
\\
\dot{\pv} &=& - \frac{\partial H'}{\partial \qv}
+ {\mathbf f}'_{\rm ext}(\qv,t). 
\ea
\end{subequations}
Eqs.~(\ref{Hamiltonian-eqn-1}) and (\ref{Hamiltonian-eqn}) are formally identical. Hence the Hamiltonian equations are covariant under redefinition of Hamiltonian and external force. 

Now the change of total Hamiltonian $H'(\qv,\pv,t)$ can be used to define another work-like quantity ${\mathcal W}$, which may be called the {\em inclusive work}:
\ba
d {\mathcal W}  &=& \frac{dH'}{dt} 
= \frac{\partial H'}{\partial \qv} \dot{\qv} 
+ \frac{\partial H'}{\partial \pv} \dot{\pv}
+ \frac{\partial H'}{\partial t} dt
\nonumber\\
&=&  {\mathbf f}'_{\rm ext}(\qv,t) \cdot d \qv
+ \frac{\partial V}{\partial t} dt
\nonumber\\
&=&  {\mathbf f}_{\rm ext}(\qv,t) \cdot d \qv
+ dV(\qv, t), 
\label{dW-cal-def}
 %=  \frac{\partial V}{\partial \lambda} \dot{\lambda}. 
\ea
where $dV(\qv, t)$ is the total differential of $V(\qv, t)$:
\be
dV(\qv, t) = \nabla_{\qv} V(\qv, t ) \cdot d \qv
 +  \frac{\partial V}{\partial t} dt. 
\ee 
 Comparing Eqs.~(\ref{dW-cal-def}) with (\ref{dW-def}), we see that the difference between two works $d {\mathcal W}$ and $d W$ is just the change of the external potential energy $dV$:
\begin{subequations}
 \label{work-relation-2}
\ba
d {\mathcal W} &=&  d W + d V, \\
{\mathcal W} &=& W  + \Delta V.
\ea
\end{subequations}

% To make distinction, we may call $W$ the {\em exclusive work} and ${\mathcal W}$ the {\em inclusive work}. 

\subsection{Time-reversals of states and of trajectories}
\label{app:T-reversal}
For any point in phase space $\xv = (\qv, \pv)$ (a microscopic state), we define its time-reversal as $\xv^* = (\qv,  - \pv)$.  We shall assume that the intrinsic Hamiltonian is invariant under time-reversal, i.e.,
\be
H(\qv, - \pv) = H(\qv, \pv). 
\ee
This implies that there is no magnetic field acting on the system.  In the presence of a magnetic field, both $\xv$ and the vector potential ${\mathbf A}$ needs to be time-reversed in order to guarantee the invariance of the intrinsic Hamiltonian.  

Now consider a trajectory in the coordinate space $\qv(\tau)$ with the time parameter $\tau$ running from $0$ to $t$, we can construct the corresponding phase space trajectory as
$\xv(\tau) = (\qv(\tau), \pv(\tau))$, where $\pv(\tau)$ may be computed from Lagrangian.  (The point here is that NOT every curve in phase space corresponding to a physical trajectory in the coordinate space!)  The time-reversal of the coordinate space trajectory can be trivially defined:
$\qv^*(\tau) = \qv(t-\tau)$.  From this, we calculate the corresponding momenta curve:
\be
\pv^*(\tau) = \left. \frac{\partial L}{\partial \dot{\qv}} \right|_{\qv = \qv(t - \tau )}
= - m \dot{\qv}(t - \tau )
= - \pv(t - \tau). 
\ee
Hence, as expected, the time-reversed trajectory in phase space is given by 
\be
\xv^*(\tau) = (\qv(t - \tau), - \pv(t - \tau)). 
\ee
Note that the initial state of the time-reversed trajectory is the time-reverse of the final state of the forward trajectory, and vice versa:
\ba
\xv^*(0) &=& (\qv(t), -\pv(t)),\\
\xv^*(t) &=& (\qv(0), -\pv(0)). 
\ea

%These results look apparently strange, because we have transformed the time variable.  

To make the Hamiltonian equations covariant, we also need to time-reverse the external force:
\be
\fv^*_{\rm ext} (\qv, \tau) = \fv_{\rm ext} (\qv, t - \tau).
\label{f-T-rev-def}  
\ee
Now the time-reversal symmetry of Hamiltonian dynamics can be formulated as follows: If $\xv(\tau) =(\qv(\tau), \pv(\tau))$ is a solution to Eqs.~(\ref{Hamiltonian-eqn}) with external force $\fv_{\rm ext} (\qv, \tau)$, the $\xv^*(\tau) =  (\qv(t-\tau ), - \pv(t - \tau ))$ is a solution  to Eqs.~(\ref{Hamiltonian-eqn}) with external force replaced by $\fv^*_{\rm ext} (\qv, t)$.  Proof is simple. 

The backward evolution operator be $U^*(t_1,t_2)$ is defined such that the system evolves from a state $\xv$ at $t_1$ to another state $\yv$ at $t_2$ under forward dynamics, if and only if it evolves from ${\yv^*}$ at $t_1$ to $\xv^*$ at $t_2$ under the backward dynamics. Mathematically we have
\begin{subequations}
\ba \yv = U(t_1, t_2) \xv  \quad 
&\longleftrightarrow& \quad 
\xv^* = U^*(t_1,t_2) \tilde{\yv},
\\
B = U(t_1,t_2) A  \quad 
&\longleftrightarrow& \quad 
A^* = U^*(t_1,t_2) B^*,  \quad\quad
\ea
\end{subequations}
where $\longleftrightarrow$ means ``if and only if''. 
%Note that the backward evolution operator also  evolves from time $t_1$ to time $t_2$, the same as the forward operator.  

\section{Derivation of Time Reversal Relation}
\label{app:PAB} 

\subsection{Derivation of Eq.~(\ref{PAB})}
Let a system starting from a Boltzmann macro-state Eq.~(\ref{rho_A-def}) at $t = 0$.  The evolved pdf at $t$ is Eq.~(\ref{rho-t-1-0}). The corresponding coarse-grained state is given in Eq.~(\ref{tilde-rho-1-0}).  Using Eq.~(\ref{rho_A-def}) we can rewrite it as 
\be  
 {\mathscr C } U(0,t) \frac{1}{|A|} \chi_A(\xv)
\equiv \sum_B{Pr}(B, t |A, 0) \, \frac{1}{|B|} \chi_B(\xv). 
\label{tilde-rho-1}
\ee 
Multiplying both sides by $\chi_{B'}(\xv)$, integrating over $\xv$, using Eqs.~(\ref{U-chi-chi}), and finally let $B'\rightarrow B$, we obtain Eq.~(\ref{PAB}). 
\ba
{Pr}(B, t |A, 0) &=& 
%\int_{B} d \omega \,   \tilde{\rho}(\xv, t)
   \frac{1}{|A|} (\chi_B, {\mathscr C } U \chi_A)
\nonumber\\
&=&  \frac{1}{|A|} |U A \cap B|. 
\label{PAB-1}
\ea
 
 As  illustrated in Fig.~\ref{Fig:partition-1.pdf}, ${Pr}(B, t |A, 0)$ is just the fraction of phase space volume of the set $ U A$ intersecting $B$.  Since $\{UA  \cap B_k, \forall k \}$ forms a partition of $UA$,
 i.e., $\cup_k UA  \cap B_k = UA$,   ${Pr}(B, t |A, 0)$ is normalized:
\be
\sum_B{Pr}(B, t|A, 0 )= 1.
\label{P-normalization}
\ee

\subsection{Derivation of Eq.~(\ref{second-law})}
Here we use the short hand $U = U(0,t)$, and $U^* = U^*(0,t)$.  The following identities can be established using definitions of operator $U$  and time reversal, as well as Liouville theorem:
\vspace{-3mm}
\begin{subequations}
\label{useful-identities}
\ba
|A| &=& |U {A}| = |U^{-1}A|,
\label{useful-identities-1}\\
U(A \cap B) &=& (U A) \cap (UB),
\label{useful-identities-2}\\
( U A)^* &=& (U^*)^{-1}A^*, 
\label{useful-identities-5}\\
( A \cap B)^* &=& A^* \cap B^*,
\label{useful-identities-4}\\
|A| &=& |A^*|. 
\label{useful-identities-3}
\ea
\end{subequations}
Using Eqs.~(\ref{useful-identities-1}), (\ref{useful-identities-2}), (\ref{useful-identities-5}),(\ref{useful-identities-4}), and (\ref{useful-identities-3}) successively, we have
\ba
 \frac {| U^*B^* \cap  A^*|}{|B^*|}
&=&  \frac {| (U^*)^{-1}(U^*B^* \cap  A^*)|}{|B^*|}
= \frac{| (U^*)^{-1}{A^*} \cap  B^*|}{|B^*|}
\nonumber\\
&=& \frac{| ( U A)^* \cap  B^*|}{|B^*|}
= \frac{|( U A \cap B )^*|}{|B^*|}
= \frac{|{ U A \cap B}|}{|{ B}|}.  
\label{PBA-1}
\ea
Substituting this back into Eq.~ (\ref{PBA}), and dividing it by Eq.~(\ref{PAB}), we obtain the Stochastic H-Theorem Eq.~(\ref{second-law}) for the transition probabilities. 

\section{Mixing and Local Mixing}
\label{app:mixing}

In ergodic theory~\cite{book-Sinai-Ergodic theory}, a measure-preserving, stationary dynamic system  is said to be {\em mixing}, if for arbitrary Boltzmann states  $A, B$:
\begin{subequations}
\be
\lim_{t \rightarrow \infty} {Pr}(B, t |A, 0) 
= \lim_{t \rightarrow \infty}
\frac{ |U(t) A \cap B| }{|A|}
= \frac{ |B|}{|\Omega(E)|},  
\label{strong-mixing-def}
\ee
where $\Omega(E) = \underset {\textit {\tiny A, E(A) =E}} {\cup} A$ is an energy shell we defined earlier.  ({Some authors use the term {\em strong mixing}, to make difference with {\em weak mixing}.}) Note that we have rewritten $U(t,0)$ as $U(t)$ because of the time-translation symmetry.  According to Eqs.~(\ref{tilde-rho-1-0}), this means that the coarse-grained evolved distribution converges to the flat distribution in the energy shell $\Omega(E)$, which is just the thermal equilibrium state:
\ba
\lim_{t \rightarrow \infty} {\mathscr C }U(t) \rho_A(\xv)
&=&  \frac{1}{|\Omega|} \sum_B \chi_B(\xv)
 = \frac{1}{\Omega(E)} \equiv \rho_{\rm EQ}(\xv).
\quad 
 \label{strong-mixing-def-2}
\ea
\end{subequations}
The time scale of this convergence is the equilibration time $\tau_{\rm EQ}$ for the entire system, which generically diverges in the thermodynamic limit. 

By contrast, our experiences tall us that the time scale over which macroscopic irreversibility emerges is independent of system size.  For dilute gases, $\tau_{\rm irrev}$ is typically the mean free time of gas particles, i.e., the time duration between two consecutive collisions experienced by a particle.  Furthermore, if a system is driven by external forces, energy is not conserved, the system never equilibrate, and the mixing property can not even be defined. Nonetheless, the entropy keeps increasing, as shown by Theorem \ref{thm-2}.  Hence the mixing property is NOT the origin of macroscopic irreversibility.  We must look elsewhere to resolve the Loschmidt paradox.  
%We must explore which property of microscopic dynamics is responsible for the monotonic increase of entropy in the experimentally relevant time scale.  

%Our general experiences show that macroscopic irreversibility emerges in a mesoscopic time scale $\tau_{\rm irrev}$, which is independent of system size. That is, we expect
%\be S_{\rm G}[\tilde{\rho}(t)] \geq S_{\rm G}[\tilde{\rho}(t')],
%\quad \forall \,\, t - t' \gg \tau_{\rm irrev}. 
%\label{second-law-Gibbs} \ee 

%For ergodic system, there is another recurrence time scale $\tau_{\rm rec}$, over which the system returns to a microscopic state that is sufficiently close to its initial state.  It scales exponentially with the system size, and is longer than the age of universe for large systems.  Hence it must be irrelevant to macroscopic irreversibility.  

Let us rewrite Eq.~(\ref{local-mixing-def-1}) into the following form:
\be
\lim_{t, \tau \rightarrow \infty}
{\mathscr C}  U(\tau,  t+ \tau)  U(0, \tau)  \rho_A  
 = \lim_{t, \tau \rightarrow \infty} {\mathscr C}
  U(\tau,  t+ \tau) {\mathscr C} U(0,\tau) \rho_A. 
\label{local-mixing-def-3}
\ee
The LHS is the state obtained if the system evolve from $\rho_A$ at $t = 0$ to time $t+ \tau$, and  is coarse-grained.  The RHS, on the other hand, corresponds to a rather different evolution history:  the system first evolves from $\rho_A$ for a time duration $\tau$, being coarse-grained at that moment, and then keep evolving for another time $t$, and finally coarse-grained again.  Hence, qualitatively speaking, local mixing means that short scale differences at time $\tau$ do not influence the macroscopic properties in the distance future $t+\tau$, as long as the system starts evolution from a Boltzmann state in a remote past $t = 0$.   Since system size is not needed in the definition of local mixing, we expect that the characteristic time scale $\tau_{\rm LM}$ of local mixing (when the limits in Eq.~(\ref{local-mixing-def-3}) converge) should be {\em independent of} system size.  
 
Local mixing is weaker than mixing.  Mixing clearly implies local mixing, since Eq.~(\ref{strong-mixing-def}) means that both terms of Eq.~(\ref{local-mixing-def-3}) reduce to $\rho_{\rm EQ}$.  On the other hand, local mixing does not guarantee ergodicity, and hence does not guarantee mixing.  Breaking down of ergodicity may happen due to two reasons: 1) there are conserved quantities other than energy, and 2) the system may exhibit spontaneous symmetry breaking in the thermodynamic limit.  In either case, the phase space $\Omega(E)$ is broken into ergodic components that are not mutually accessible.  Local mixing is however a local property of dynamics, and is insensitive to these issues.

%\vspace{2mm}
\section{Derivation of Fluctuation Theorems using Theorem \ref{thm-1}}
\label{app:FTs}
%\section{Generalization to Open Systems}
 
 We shall consider open systems, which are in contact with a heat bath with temperature $T = 1/\beta$.  To simplify the discussion, we shall make the standard assumptions that interaction between the system and  bath is infinitesimal, that the heat bath remains in thermal equilibrium all the time, and that there is no statistical correlation between the system and heat bath.  % Relaxing of these assumptions does not change our essential results, though makes the derivation technically more involved. 

Let us define a new concept, {\em Boltzmann free energy}:
%~\footnote{Equation (\ref{Boltzmann-F-def}) appears  similar to free energy functionals widely used in {\em density functional theory} (DFT).  There is however an  important difference: in Eq.~(\ref{Boltzmann-F-def}), $\rho_A(\xv)$ is a coarse-grained pdf in high dimensional phase space, whereas in DFT $\rho(\xv)$ is usually a one-particle pdf in three dimensional space.   }:
\ba
\vspace{-2mm}
{\mathcal F}_{\rm B}(A, T)&=& \int d \omega \,
 \rho_A(\xv) \Big( 
H(\xv)  + T\, \log \rho_A(\xv) \Big)
\nonumber\\
&=& E(A) - T S_{\rm B}(A),
 \label{Boltzmann-F-def}
\ea
where $E(A) \equiv \int d \omega \, \rho_A(\xv) H(\xv)$ is the energy of macro-state $A$,  and $H(\xv)$ does not include the influences of external forces.  ({The variance of energy inside $A$ is negligible, since $H(\xv)$ is one of the macroscopic quantities chosen to define partition.  For details, see App.~\ref{app:Math}.}) For a coarse-grained Gibbs state, as given in Eq.~(\ref{Pr(A)lpha-def}), we can also define the {\em coarse-grained free energy}:
%\begin{subequations}
\ba
%\vspace{-2mm}
F[\tilde{\rho}, T]  &\equiv&
  \int d \omega \,  \tilde{\rho}(\xv) \left(  H 
+ T\,\log \tilde{\rho} (\xv) \right) 
\nonumber\\
&=& \sum_A p(A) \big( {\mathcal F}_{\rm B}(A, T) 
+  T \log p(A) \big),
 \nonumber
%\label{F-tilde-def}
\label{F-tilde-def-0}
%\vspace{-2mm}
\ea
%\end{subequations} 
which reduces to the usual equilibrium free energy if $\tilde{\rho}$ is the equilibrium Gibbs distribution.  
These free energies are applicable for arbitrary non-equilibrium systems.  
%Boltzmann free energy and coarse-grained free energy are the counterparts of Boltzmann entropy and coarse-grained Gibbs entropy for open systems. 

%\label{sec:thermostated}
% Note that Boltzmann free-energy is applicable for arbitrary Boltzmann states, which may be far away from equilibrium.  %which will become clear below as we generalize theorems 1 and 2 to thermostated systems. 

%For technical details, see App.~\ref{app:work}.

% Here ${\mathcal F}_{\rm B}(A, T)$ is a very general and flexible concept, applicable for arbitrary non-equilibrium system.  We shall call it  {\em Boltzmann free energy}, in echo of its essential connection with Boltzmann's notions about macro-state and entropy.  

%{\bf Work and Dissipation}   \quad 
%\label{sec:thermo-stat}

Consider a process where the system starts from a Boltzmann state $A$ and arrives at state $B$, at the same time the external force do work ${W}$.  Let $Q$ be the energy transferred from the system to the heat bath (in the form of heat).   The entropy of heat bath increases by $\beta Q$.  The change of total energy for the combined system is  the total work done by the external force: $W =  Q + E(B) - E(A)$.  The total change of entropy is then
\ba
\Delta S^{\rm tot} 
&= &  S_{\rm B}(B) - S_{\rm B}(A)  + \beta Q
\label{Delta-S-1}\\
&=& \beta {W} - \beta 
\left( {\mathcal F}_{\rm B}(B, T) - {\mathcal F}_{\rm B}(A, T) \right). 
\nonumber
\ea
Substituting this back to Eq.~(\ref{second-law}), we obtain the {\em Stochastic H-Theorem for open systems}:
\ba
&&  e^{ - \beta  {W}  - \beta  {\mathcal F}_{\rm B}(A, T)} 
 {{Pr}( B, t | A, 0 ; {W})}
\nonumber\\
&=& e^{ - \beta {\mathcal F}_{\rm B}(B, T)} 
{{{Pr}^*}(A^*,t|B^*,0; -{W})}.
\label{second-law-open}
\ea
Here $ {{Pr}( B, t | A, 0 ; {W})}$ should be understood as the joint probability that the system transits from Boltzmann states $A$ to $B$, while the external forces do work $W$.  Below, we shall use Eq.~(\ref{second-law-open}) to derive the Fluctuation Theorems due to Crooks, Evans and Searles, as well as the work identities due to Jarzynski~\cite{Jarzynski-1997}, and due to Bochkov and Kuzovlev~\cite{BK-1977}.

The equilibrium Helmholtz free energy (with respect to the unperturbed Hamiltonian $H$) $F(T)$ is given by the usual form of Gibbs factor integrated over the entire phase space: 
\ba
e^{- \beta F(T)} &=& 
\int d \omega \, e^{- \beta E(\xv)}
%=  \sum_A |A| \, e^{- \beta E(A)}
%\nonumber\\
%= \sum_A e^{-\beta {\mathcal F}_{\rm B}(A, T)}.
%= \sum_E \Omega(E) \, e^{- \beta E}, 
\label{Helmholz-F-def}
\ea
Using the phase space partition, we can express the above result as the sum of  Boltzmann free energy ${\mathcal F}_{\rm B}(A, T) $ over Boltzmann states $A$:
\be
e^{- \beta F(T)} = \sum_A \int_A d \omega \, e^{- \beta E(\xv)}
=   \sum_A |A| \, e^{- \beta E(A)}
= \sum_A e^{-\beta {\mathcal F}_{\rm B}(A, T)}. 
%= \sum_E \Omega(E) \, e^{- \beta E}.
\ee
This result indicates that thermal equilibrium is a special macro-state with probability of Boltzmann states given by:
\be
p_{\rm EQ}(A) = e^{\beta (F - {\mathcal F}_{\rm B} (A))}. 
\label{p_EQ-A}
\ee  
Indeed, one may easily verify that the coarse-grained free energy Eq.~(\ref{F-tilde-def-0}) reduces to the equilibrium free energy if $p(A)$ is replaced by Eq.~(\ref{p_EQ-A}).  

If the system is perturbed by an external potential $V(\qv)$, it will settle down to a modified equilibrium state with total Hamiltonian $H'= H + V$.  It is convenient to choose $V(\qv)$ to be one of the macroscopic quantities in the set $\{{\mathscr M}_k\}$, and let $V(A)$ be the mean value of $V(\xv)$ in the cell $A$, defined in Eq.~(\ref{M-cell-approx}).   The corresponding equilibrium free energy $F'$ is then given by
\ba
e^{-\beta F'} &=& \int d \omega \, e^{-\beta H - \beta V} 
= \sum_A |A| \, e^{-\beta E(A) - \beta V(A)} 
\nonumber\\
&=& \sum_A e^{- \beta {\mathcal F}_{\rm B} (A) - \beta V(A)}. 
\label{F'-def}
\ea
Hence in the modified equilibrium state, the probability of the Boltzmann state $A$ is 
\be
p'_{\rm EQ}(A) = e^{\beta (F' - {\mathcal F}_{\rm B} (A) - V(A))}. 
\label{p_EQ-A-mod}
\ee  

% From the Stochastic H-Theorem for open systems Eq.~(\ref{second-law-open}),  we can derive a well-known identity due to Crooks, which relates the distribution of entropy production in a forward process to that of a backward process.  These two processes start from different thermal equilibrium.  

Let us now consider a system initially at $t=0$ in thermal equilibrium with an intrinsic Hamiltonian $H$.  We then apply an external driving force, which is generically non-conservative, and drive it to some non-equilibrium state at $t = \tau$.   We shall use Eq.~(\ref{f-decomp}) to decompose the external force into a conservative part  and another non-conservative part.  Furthermore,we shall choose $V(\qv,0) = 0$ at the initial time $t= 0$.  Both $H(\qv,\pv)$ and $V(\qv,t)$ are invariant under time-reversal, $V(\qv^*, t) = V(\qv, t), H(\xv^*) = H(\xv)$.  Consequently we have 
\ba
V(A^*, t) = V(A, t), \quad 
{\mathcal F}_{\rm B}(A^*) = {\mathcal F}_{\rm B}({A}). 
\label{V-F-t-rev}
\ea

Let us now multiply both sides of Eq.~(\ref{second-law-open}) by $e^{\beta (F - V(B, t))}$, sum over $A, B$, and use Eq.~(\ref{V-F-t-rev}), we obtain 
\ba
&&  \sum_{A,B} e^{ - \beta ( { W} + V(B, t) )} e^{\beta ( F -  {\mathcal F}_{\rm B}(A)) } 
{ Pr}(B, t | A, 0; { W}) 
\label{detailed-balance-3}\\
&=& \sum_{A,B}  e^{\beta ( F - F')} 
e^{ \beta ( F'- {\mathcal F}_{\rm B}(B^*) - V(B^*,t)) } 
{{Pr}^*}(A^*,t | B^*,0; - { W}), 
\nonumber
\ea
 where $F'$ is defined in Eq.~(\ref{F'-def}), with $V(A) = V(A,t)$.   Now, according to Eq.~(\ref{p_EQ-A}), in the initial state the system is in Boltzmann state $A$ with probability $e^{\beta (F - {\mathcal F}_{\rm B} (A))}$.  On the other hand, given that the system starts from $A$, the probability that the system evolves to $B$, and at the same time the external forces do work $W$ is given by ${ Pr}(B, t | A, 0; { W})$.  Note that $W$, defined as the change of intrinsic Hamiltonian, is the exclusive work, whilst the inclusive work ${\mathcal W}$ is related to $W$ via Eq.~(\ref{work-relation-2}), which according to our present setting, becomes
 \be
 {\mathcal W} =   W  +V(B, t) - V(B, 0) =  W  +V(B, t),
 \ee
 since we have chosen $V(\qv,  0) = 0$.   For definitions of exclusive work and inclusive work, see App.~\ref{app:work}.  Let us further multiply both sides of Eq.~(\ref{detailed-balance-3}) by $\delta ({\mathcal W} -  W - V(B, t))$, and sum over $W$ (It really should be integral over $W$, if $W$ is treated as a continuous variable.  However notation does not matter here.).  The LHS then becomes 
\ba
\sum_{W, A, B}  \delta ({\mathcal W} -  W - V(B, t))
e^{ - \beta ( { W} + V(B, t) )}
 e^{\beta ( F -  {\mathcal F}_{\rm B}(A)) } 
{ Pr}(B, t | A, 0; { W}) . 
\ea
But this is just $e^{- \beta {\mathcal W}}$ multiplying the probability density of the inclusive work done by the external forces, $p_F({\mathcal W})$.  The RHS of Eq.~(\ref{detailed-balance-3}), on the other hand, becomes
\ba
&& e^{\beta ( F - F')} 
\sum_{W, A, B}  \delta ({\mathcal W} -  W - V(B, t))
e^{ \beta ( F'- {\mathcal F}_{\rm B}(B^*) - V(B^*, t)) } 
{{Pr}^*}(A^*,t | B^*,0; - { W}) \nonumber\\
&=& e^{\beta ( F - F')} 
\sum_{W, A, B}  \delta ({\mathcal W} -  W - V(B, t))
e^{ \beta ( F'- {\mathcal F}_{\rm B}(B) - V(B, t)) } 
{{Pr}^*}(A,t | B,0; - { W}),
\label{RHS-crooks-prep}
\ea
where we have changed the running variables $A^* \rightarrow A, B^* \rightarrow B$, and have used 
Eq.~(\ref{V-F-t-rev}).  

We shall now define the {\em backward process}, where the system starts (at time $t =0$) from thermal equilibrium with Hamiltonian $H + V(\qv, t)$, and evolves according to the time-reversed external force (defined in Eq.~(\ref{f-T-rev-def})), to some other non-equilibrium state at $t$.  
 The system is then initially in a Boltzmann state $B$ with probability $e^{\beta F' -\beta {\mathcal F}_{\rm B}(B) -\beta V(B, t)}$, where $F'$ is defined in Eq.~(\ref{F'-def}).  Now, if the system evolves from $B$ to $A$, and at the same time the exclusive work is $-W$, as shown in Eq.~(\ref{RHS-crooks-prep}), then inclusive work done by the external forces is then $- W + (V(A^*, 0) - V(B^*, t)) = - W - V(B^*, t) = - {\mathcal W}$.  (Recall that the external potential is also time-reversed. )  Hence the summation in Eq.~(\ref{RHS-crooks-prep}) is just the probability density of the inclusive work in the backward process, evaluated at $- {\mathcal W}$: $p_R(-  {\mathcal W})$.  Equating the preceding two results, we obtain  the well-known {\em Crooks Fluctuation Theorem}~\cite{Crooks-PRE-1999}: 
\be
p_F( {\mathcal W}) e^{- \beta  {\mathcal W}} 
= p_R(-  {\mathcal W}) \, e^{\beta (F - F')}. 
\label{Crooks-FT}
\ee  
Note that the work $ {\mathcal W}$ appearing here is the {\em inclusive work},  defined as the difference of the {\em total Hamiltonian} $H(\qv, \pv) + V(\qv, t)$.   Note also that the forward and backward processes start from different equilibrium states, with equilibrium free energies $F, F'$ respectively.

Summing Eq.~(\ref{Crooks-FT}) over ${\mathcal W}$, and using the renormalization condition $\sum_{\mathcal W} p_R(-  {\mathcal W})  =1$, we obtain the famous {\em Jarzynski equality}~\cite{Jarzynski-1997}:
\be
\left\langle e^{- \beta {\mathcal W}} \right\rangle
=  e^{- \beta ( F'  - F)}
=  e^{- \beta \Delta F}.  
\label{Jarzynski}
\ee
where $\langle \,\, \cdot \,\, \rangle$ means averaging over the forward process.   Note that Eq.~(\ref{Jarzynski}) is valid regardless of whether the external force is non-conservative nor not.  It is also independent of the choice of the conservative potential $V(\qv,t)$, as long as the free energy $F'$ is defined accordingly by Eq.~(\ref{F'-def}).   Hence Jarzynski equality is more general than what is allowed in the original proof by Jarzynski himself.

If we {\em choose}  $V(\qv,t) = 0$, then $F' = F$, and ${\mathcal W} = W$. Hence the forward and backward processes start from the same equilibrium state.  The Crooks FT becomes
\be
p_F( { W}) e^{- \beta  { W}} 
= p_R(-  { W}).  
\label{Crooks-FT-1}
\ee
Note, however, since the external force is generically different for the forward and backward processes, the probability densities of work for the forward and backward processes are also generically different, $p_F(W) \neq p_R(W)$.  Nonetheless, if we sum this relation over $W$, RHS reduces to unity because of renormalization, and we obtain the {\em Bochkov-Kuzovlev equality} (BKE)\cite{BK-1977}:
\be
\left\langle e^{- \beta { W}} \right\rangle = 1.
\label{Bochkov-Kuzovlev-eq}
\ee
Same as JE, BKE is applicable for arbitrary process starting from thermal equilibrium. Note however ${ W}$ is exclusive work, defined as the change of intrinsic Hamiltonian $H(q,p)$.  Using Eq.~(\ref{work-relation-2}), we can also express BKE in terms of the inclusive work ${\mathcal W}$ as:
\be
\left\langle e^{- \beta {\mathcal W} + \beta \Delta V} \right\rangle = 1. 
\ee
This is NOT the same as JE, Eq.~(\ref{Jarzynski}).  Hence the informations encoded in BKE and JE are {\em physically different}, even though both of them are valid for arbitrary non-equilibrium processes starting from equilibrium.  Their differences vanish only if we choose $\Delta V = 0$.

\begin{figure}[thp!]
\begin{center}
%\subfigure[]{	}
		\includegraphics[width=8cm]{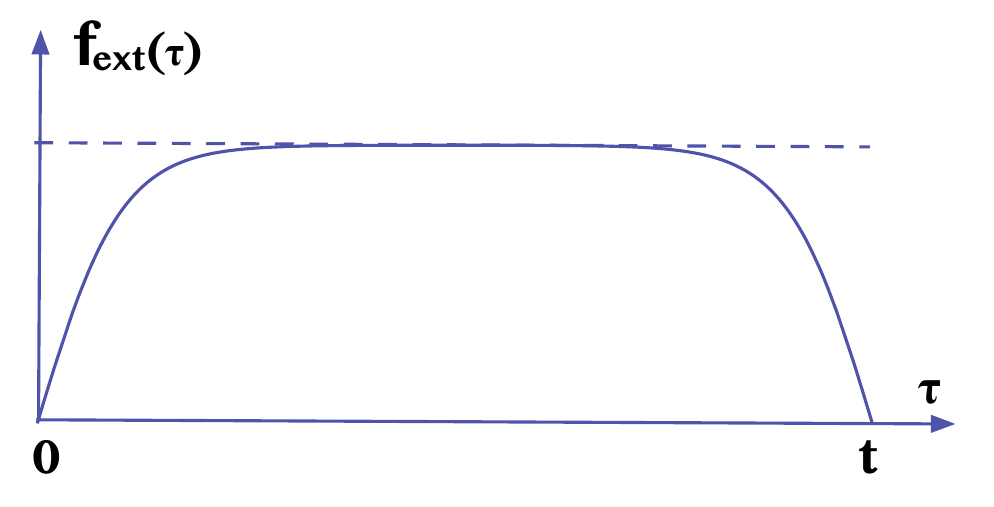}
\caption{A special case where the external force is turned on at $t = 0$ and turned off at $t$ symmetrically. }
\label{Fig:f_ext-t}
\end{center}
\vspace{-5mm}
\end{figure}

Now let us consider a special case where the external force is turned on at $t = 0$ and turned off at $t = \tau$, so that ${\bf f}_{\rm ext}(\qv, 0)  =  {\bf f}_{\rm ext}(\qv, t) = 0 $.  Furthermore, we shall assume that the turning-on and turning-off are symmetric: 
\be
{\bf f}_{\rm ext}(\qv, \tau) =  {\bf f}_{\rm ext}(\qv,  t - \tau )
= {\bf f}^*_{\rm ext}(\qv, \tau). 
\ee
where in the last equality we have used Eq.~(\ref{f-T-rev-def}).  Hence the external force, and consequently also the evolution operator are the same for the forward and backward processes.  Furthermore, we {\em choose} $V (\qv, t) = 0$, so that the forward and backward processes start with the same equilibrium states, and that the probability densities for work $p_F({\mathcal W})$ and $p_R({\mathcal W})$ also become identical.  The relation Eq.~(\ref{Crooks-FT-1}) then becomes the {\em Evans-Searles Fluctuation Theorem}~\cite{Sevick-ARPC-2008}:
\be
e^{ - \beta W}  {p({ W})} = {p(- { W})}.  
\label{Evans-Searles-FT}
\ee 

% In Appendix \ref{app:BKE-JE} we derive from these results the famous Bochkov-Kuzovlev equality and Jarzynski equality.  

\section{Evolution of Free Energy for Open Systems}
\label{app:Open systems}

Let us now generalize Theorem \ref{thm-2} to open systems.  The combined system starts evolution from  a product state: $A \otimes \Omega^b(E^b)$, with total entropy $S_{\rm B}(A) + S_{\rm B}^{\rm b}(E^b)$.  No measurement is performed on the system.  At time $t$, the combined system has probability $p(B, { W})$ in state $B \otimes \Omega^b(  E^b + \delta E^b)$~\footnote{Note that $p(B,W)$ depends on time $t$, even though we do not explicitly show it. }, where $\delta E^b =  E(A) +{ W} - E(B)$, with ${ W} $ the work done by the external forces.   The coarse-grained entropy of the combined system in the final state is
\ba
S_{\rm G}^{\rm tot}(t)
&=& \sum_B p(B,W) 
%\nonumber\\
%&\times& 
\log \frac{|B| 
|\Omega^b(E^b + \delta E^b )|}{p(B,W)}  \\
%&=&  \sum_B p(B) \log \frac{|B|}{p(B)}
%+ \sum_B p(B) \log |\Omega^b(E^{\rm tot} -  E(B))|
%\nonumber\\
%&=& \sum_A p(A) \log \frac{|A|}{p(A)}
%- \beta \sum_A p(A)  E(A)
%\nonumber \\
%&& + \log |\Omega^b(E^{\rm tot})|
%\nonumber \\
%\label{S-tot-t-open}
%:
&=& S^b_{\rm B}(E^{\rm tot}) 
 + \beta \Big( E(A) + \langle { W} \rangle_t 
 -  F[\tilde{\rho}(t),T] \Big), \nonumber 
\ea
where in the second equality, we have done some manipulations similar to those leading to Eq.~(\ref{Delta-S-1}).  $F[\tilde{\rho},T]$ is defined in Eq.~(\ref{F-tilde-def-0}), while $ \langle { W} \rangle_t  = \sum_B\sum_W p(B,W)W $ is the average work done by the external force up to time $t$.  According to Theorem \ref{thm-2}, this increases monotonically in long time scales, which means for $t, \tau \gg \tau_{\rm LM}$ we have 
\ba
&& \langle { W} \rangle_{t + \tau} - F[\tilde{\rho}(t + \tau),T ] 
+  F[\tilde{\rho}(0),T ] 
\nonumber\\
&\geq& \langle { W} \rangle_{t} - F[\tilde{\rho}(t),T ]  +  F[\tilde{\rho}(0),T ] 
\geq 0. 
\ea
From this we obtain the following two theorems:
\begin{theorem} 
\label{maximal-work} ({Principle of Maximal Work}) 
For thermostated systems driven by external forces, the average work done by the external forces is greater than the change of coarse-grained free energy, and the difference increases with time. 
\end{theorem}

\begin{theorem} 
\label{minimal-F} (Principle of Minimal Free Energy) 
For non-driven thermostated systems, the coarse-grained free energy $F[\tilde{\rho}(t), T]$ monotonically decreases with time.  
\end{theorem}

Note that the free energy involved here is the coarse-grained free energy, which is a property of the instantaneous non-equilibrium macro-state $\tilde{\rho}(t)$.  
%Theorems \ref{maximal-work} and \ref{minimal-F} are therefore different from the work relations derived by  Jarzynski, and by Bochkov and Kuzovlev, where the free energy always refer to equilibrium states.  

%Together these results resolve the fundamental conflict between the second law and Liouville theorem, and provide a fully consistent explanation for the emergence of macroscopic irreversibility from microscopic reversible dynamics. They constitute a solid foundation for non-equilibrium classical statistical mechanics. I believe that these results

\end{document}